\documentclass[]{aa} 
\usepackage{graphicx}
\usepackage{epsfig}
\usepackage{lscape} 
\def\gs{{_>\atop^{\sim}}}   
\def\cgs{ ${\rm erg~cm}^{-2}~{\rm s}^{-1}$ }    
   
\begin{document}   
\title{The HELLAS2XMM survey: VIII. Optical identifications    
of the extended sample.   
\footnote{Based on observations collected at the    
European Southern Observatory, Prog. ID 67.A-0401, 68.A-0514,    
69.A-0563 and 072.A-0633,    
and at the Telescopio Nazionale Galileo, Prog. ID 1\_15\_083.    
Based also on observations made with XMM-Newton, an ESA science mission.}}   
   
\author{F. Cocchia\inst{1,2,3}, F. Fiore\inst{1}, C. Vignali\inst{4,5}  
M. Mignoli\inst{5}, M. Brusa\inst{6},   
A. Comastri\inst{5}, C. Feruglio\inst{1,2},    
A. Baldi\inst{7}, N. Carangelo\inst{8}, P. Ciliegi\inst{5},    
V. D'Elia\inst{1}, F. La Franca\inst{9}, R. Maiolino\inst{10},    
G. Matt\inst{9}, S. Molendi\inst{8}, G.C. Perola\inst{9},    
S. Puccetti\inst{1,11}}   
       
\institute {INAF-Osservatorio Astronomico di Roma \\   
via Frascati 33, Monteporzio-Catone (RM), I-00040 Italy   
\and   
Dipartimento di Fisica, Universit\`a di Roma Tor Vergata\\ 
Via della Ricerca Scientifica 1, I-00133 Rome, Italy  
\and   
INAF--Osservatorio Astronomico di Brera   
via Brera 28, I--20121 Milano, Italy. 
\email{cocchia@brera.mi.astro.it} 
\and   
Dipartimento di Astronomia, Universit\`a di Bologna  \\ 
via Ranzani 1, I-40127 Bologna, Italy  
\and   
INAF--Osservatorio Astronomico di Bologna  \\ 
via Ranzani 1, I-40127 Bologna, Italy  
\and   
Max Planck Instit\"ut f\"ur Extraterrestrische Physik (MPE)\\ 
Giessenbachstr. 1, D--85748 Garching, Germany  
\and   
Harvard-Smithsonian Center for Astrophysics (CfA)\\ 
60 Garden str, Cambridge 02138 MA, USA  
\and   
INAF--IASF \\ 
via Bassini 15, I-20133 Milano, Italy 
\and   
Dipartimento di Fisica, Universit\`a Roma Tre  \\ 
via della Vasca Navale 84, I--00146 Roma, Italy 
\and   
INAF--Osservatorio Astrofisico di Arcetri \\  
Largo Enrico Fermi 5, I-50125 Firenze, Italy 
\and   
ASI Science Data Center, ASDC c/o ESRIN\\ 
 via G. Galilei, 00044 Frascati,  Italy  
}        
\date{October, 10 2006}   
 
   
\abstract 
{} 
{Hard X-ray, large-area surveys are a fundamental complement  
of ultra-deep, pencil-beam surveys in obtaining a more complete  
coverage of the AGN luminosity--redshift plane and finding  
sizeable samples of ``rare'' AGN.} 
{We present the results of the photometric and spectroscopic 
identification of 110 hard X-ray selected sources from 5 additional 
XMM--{\it Newton} fields, nearly doubling the original HELLAS2XMM 
sample.  Their 2--10 keV fluxes cover the range 
$6\times10^{-15}-4\times10^{-13}$ \cgs and the  total area 
surveyed is $\sim 0.5$ deg$^2$ at the bright flux limit.  We 
spectroscopically identified 59 new sources, bringing the 
spectroscopic completeness of the full HELLAS2XMM sample to almost 
70\% over a total area of $\sim1.4$ deg$^2$  at the bright flux 
limit.  We found optical counterparts for 214 out of the 232 X-ray 
sources of the full sample down to $R\sim25$. We measure the flux  
and luminosity of the [OIII]$\lambda5007$ 
emission line for 59 such sources.} 
{Assuming tha most high X--ray-to-optical flux ratio sources are 
obscured QSOs, we use the full HELLAS2XMM sample and the CDF 
samples, to estimate their logN--logS.  We find an obscured QSO 
surface densities of 50$\pm$23 and 100--400 deg$^{-2}$ down to flux 
limits of $10^{-14}$ and $10^{-15}$ \cgs, respectively.  At these flux 
limits the fraction of X--ray selected obscured QSO turns out to be 
similar to that of unobscured QSO. Since X--ray selection misses most 
Compton thick AGN, the number of obscured QSO may well outnumber that 
of unobscured QSOs. 

We find that hard X--ray selected AGNs with detected [OIII] emission 
span a wide range of $L_{2-10 keV}/L_{[OIII]}$ with a logarithmic 
median of (2.14$\pm$0.38).  This is marginally higher than that of a sample of 
optically selected AGNs (median 1.69 and
interquatile range 0.30), suggesting that optically 
selected samples are at least partly incomplete, and$/$or [OIII] emission  
is not a perfect isotropic indicator of the nuclear power.  
The seven X--ray Bright, Optically Normal Galaxy (XBONG) candidates in 
the sample have $L_{2-10 keV}/L_{[OIII]}\gs1000$, while their X-ray 
and optical luminosities and obscuring column density are similar to 
those of narrow-line AGNs in the same redshift interval 
(0.075--0.32). This suggests that while the central engine of 
narrow-line AGNs and XBONGs looks similar, the narrow-line region in 
XBONGs could be strongly inhibited or obscured. 
} 
{}   
 

\authorrunning {Cocchia et al.}   
\titlerunning {HELLAS2XMM optical identifications}   
   
\maketitle   
   
\section{Introduction}   
   
Deep {\it Chandra} and XMM--{\it Newton} hard X--ray surveys have been 
able to detect the sources making the majority of the Cosmic X-ray 
Background (XRB) below 6--7 keV (Giacconi et al. 2002, Bauer et al. 2004,  
Moretti et al. 2003, Worsley et al. 2004, Brandt \& Hasinger 
2005). However, deep surveys cover only a fraction of square degree of 
sky making difficult to find sizable samples of medium- and high-luminosity sources.  
To obtain a more complete coverage of the redshift-luminosity plane,  
complementing deep surveys, and to compute 
an accurate luminosity function over wide luminosity and redshift 
intervals, a much larger area, of the order of a few square degrees, 
needs to be covered.  Furthermore, large-area surveys can also provide 
sizeable samples of ``rare'' objects. To these purposes, we are carrying 
out the HELLAS2XMM serendipitous survey using suitable XMM--{\it 
Newton} archival observations (Baldi et al. 2002). As a first step, we 
have presented in Fiore et al. (2003) the optical identification of 
122 hard X-ray selected sources detected in five XMM--{\it Newton} 
fields (hereafter the 1dF sample). Here we report the results of the photometric and 
spectroscopic identification of 110 X-ray sources from five additional 
XMM--{\it Newton} fields,  hereafter the HELLAS2XMM second source sample,  
nearly doubling the original HELLAS2XMM 
sample and bringing the total area surveyed to about 1.4 deg$^2$   
 at a flux limit of $\sim 10^{-13}$\cgs. 
   
One of the most interesting findings of the optical identifications of 
the HELLAS2XMM sample, as well as of other hard X--ray selected {\it 
Chandra} and XMM--{\it Newton} surveys (see, e.g., ChaMP: Silverman et 
al. 2005; SEXSI: Eckart et al. 2005; CLASXS, CDFN, CDFS: Barger et 
al. 2005; CDFS, CDFN: Giacconi et al. 2002, Hasinger et al. 2003; 
XMM/2dF Georgakakis et al. 2004;  HBS: Della Ceca et al. 2004) is the discovery that AGN activity 
spans a range of  
properties much  
 wider than it was thought based on optically and soft X--ray selected AGNs.  
 Indeed, hard X--ray selection provides  a more 
complete view of AGN activity, being more efficient in  
selecting sources possibly missed by  selection in other bands. For example,  
obscured and low--luminosity AGN several of which are left out 
 in optical or soft X--ray selected samples. 
 While most of optically 
and soft X--ray selected AGNs have an X--ray-to-optical (R$-$band)  
flux ratio X/O 
\footnote{$X/O=\log{\frac{f_X}{f_R}}=\log{f_X}+\frac{R}{2.5}+C$, where C depends on  
the chosen X--ray band and optical filter.} 
between 0.1 and 10 (see, e.g., Maccacaro et al. 1988, Laor et al. 1997,  
Hasinger et al. 1998, Mineo et al. 2000), XMM--{\it Newton} and 
{\it Chandra} hard X--ray surveys have  selected AGNs with X--ray-to-optical  
flux ratio well outside this range.  In particular, the 
HELLAS2XMM survey has detected both a population of relatively 
bright X--ray sources with faint optical counterparts (and thus 
characterized by high X/O values; Fiore et al. 2003) and several 
X--ray sources in otherwise inactive and optically bright galaxies 
(with relatively low X/O ratios; named X--ray bright, optically normal 
galaxies, XBONGs, Fiore et al. 2000, Comastri et al. 2002). 
   
The former population includes about 15--20\% of the sources selected 
in the 2--10 keV band to have X/O more than one order of magnitude 
higher than that of typical broad-line, type 1 AGNs. Based on optical 
spectroscopy (Fiore et al. 2003), optical-to-near--infrared 
colors (Mignoli et al. 2004), near-infrared (Maiolino et al. 2006) and 
X--ray spectroscopy (Perola et al. 2004), we concluded that the 
majority of the HELLAS2XMM sources with X/O$\gs10$ are  optically obscured QSOs at 
z$\gs1$.  At fainter X--ray and optical fluxes, the situation is less 
clear, mainly because most of the optical counterparts of the high X/O 
sources are unaccessible to optical spectroscopy even with 10m-class 
telescopes.   
 
The XBONGs, on the other hand,  
are found at low redshift ($\sim$ 0.1--0.3), with X--ray luminosities 
between $10^{42}$ and $10^{43}$ erg s$^{-1}$, bright optical 
counterparts (R$<21$), thus relatively low X/O, and absorption--dominated  
optical spectra without strong nuclear emission (Comastri et 
al. 2002). It has been suggested that XBONGs selected in {\it Chandra} 
and XMM--{\it Newton} surveys are actually typical AGNs classified as 
normal galaxies just because of some observational biases (dilution of 
the nuclear spectrum by the host galaxy, inadequate set--up for 
optical spectroscopy in terms of wavelength range covered, signal-to-noise  
ratio, spatial resolution, see e.g.  Moran et al. 2002, 
Severgnini et al. 2003, Georgantopoulos \& Georgakakis 2005).  
  
In this paper we present the optical identification follow--up 
observations of five additional fields of the  HELLAS2XMM survey 
and focus the discussion on both the high X/O sources and the XBONGs, 
providing quantitative constraints to the space density of highly 
 optically obscured, high-luminosity  QSOs and proposing a quantitative  
definition of XBONGs.  The paper is organized as follows: Sections 2 
and 3 present the results of the optical photometric and spectroscopic 
identifications of the HELLAS2XMM second source sample; Section 4 
discusses our main findings; Section 5 reports our conclusions.  A 
$H_0=70$ km s$^{-1}$ Mpc$^{-1}$, $\Omega_M$=0.3, 
$\Omega_{\Lambda}=0.7$ cosmology is adopted throughout.

\section{Optical identifications}   
   
The HELLAS2XMM second source sample includes sources detected in the 
2--10 keV band in five XMM--{\it Newton} fields: A1835, 
IRAS13349$+$2438, GD153, Mrk421, BPM16274 (see Baldi et al. 2002 for 
details on the observations and X--ray data reduction).  We have 
obtained relatively deep (exposure time 3--15m per image, limiting 
magnitude R=24--25) optical images for the majority of the sources in 
these five fields using EFOSC2 at the ESO 3.6m telescope, FORS1 at VLT 
and DOLORES at the TNG.\\ We exclude from the following analysis those 
areas, in the five fields, not covered by R$-$band images. This, 
together with the fact that Mrk421 field has been observed in 
window mode and that a large-area centered on the A1835 cluster of 
galaxies has been excluded from the analysis,  
 reduces the total area 
surveyed to about 0.5 deg$^2$.  The total number of hard X--ray 
selected sources making the HELLAS2XMM second source sample is 
110, bringing the total number of sources in the full HELLAS2XMM 
sample  with optical coverage to 232 in ten XMM--{\it Newton}  
fields.

Optical images were bias subtracted, flat--field divided, and flux 
calibrated using observations of standard stars acquired during each 
night.  Optical and X-ray images were brought to a common astrometric 
reference frame using bright AGNs (from 5 to 15 AGNs per 
field). Typical systematic shifts were of the order of 1$''$, the 
maximum shift was of $\sim2''$. 
    
Source detection in the optical images was performed using the 
SExtractor package (Bertin et al. 1996) and we  visually searched for optical 
counterparts of the X--ray selected sources 
  within a conservative matching radius of 6 arcsec from the astrometrically  
corrected X­-ray centroid.  
For X-ray sources with multiple optical counterparts we provide, in Table 1, 
the probability P 
of chance coincidence\footnote{$P=e^{-\pi\,r^2\,n(R)}$ where r is the X--ray--optical displacement 
and n(R) the number counts of galaxies as a function of their magnitude from  
Metcalfe et al. (2001)} (see also Brusa et al. 2003). 
   
Optical counterparts brighter than R$\sim24.5$ within $\sim6''$ from 
the X--ray position were found for 103 sources.  For 15 of them there 
are two candidate counterparts inside the X--ray error--box.  Table 1 
gives for each source the X--ray position, the position of the most 
likely optical counterpart (identified  on the basis of its magnitude, 
displacement from the X-ray position and probability  
of chance coincidence, classification of the optical 
spectrum), the displacement between the X--ray and optical positions, 
the X--ray flux and the R magnitude (or lower limit) of the optical 
counterpart. In three cases, more than 
one optical source might contribute to the detected X-ray emission  
(see notes to Table 1). 
  
\begin{figure}   
\centering    
\includegraphics[width=8cm]{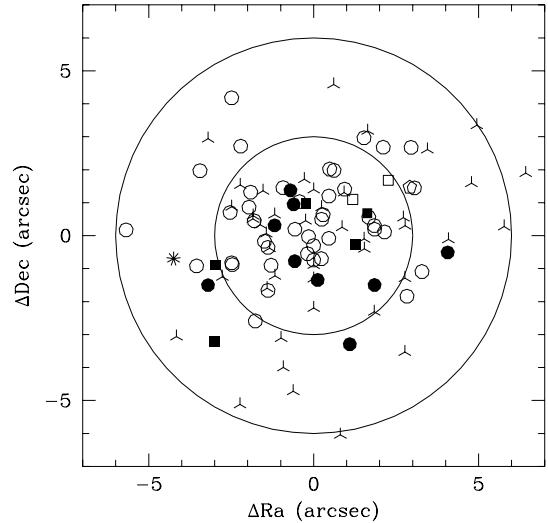}   
\caption{The displacement between the X-ray position and the position  
of the nearest optical counterpart for the HELLAS2XMM  second source sample. 
 Open circles = broad-line AGN; filled circles = narrow-line  
AGN; filled squares = emission-line galaxies; open squares = early-type  
galaxies; stars = stars; pentagons = groups/clusters of galaxies; skeleton  
triangles = unidentified objects. The two circles have radii of 3 and  
6 arcseconds, respectively.}  
\label{radec}   
\end{figure}

The average displacement between the X--ray centroid and the position 
of the most likely optical counterparts is of $2.4''\pm1.4''$.  For 
$\sim67$\% of the sources the displacement is $<3''$ (see figure 
\ref{radec}). These shifts are slightly worse than those found for the 
HELLAS2XMM  1dF sample  
(average displacement of $2.0''\pm1.4''$ and 
80\% of the sources within $3''$, Fiore et al. 2003), most likely 
because the average off-axis angle (and therefore the average Point 
Spread Function) of the sources in the second sample is higher than 
for the sources in the 1dF sample, due to the exclusion of the central 
parts of two fields (Mrk421 and A1835).

\section{Optical spectroscopic redshifts and classification}   
 
 Our main effort was that of starting to 
explore the region of high X$/$O values to 
obtain an optical spectroscopy as uniform as possible 
regarding the X--ray to optical flux ratio. This has been roughly 
achived up to $X/O$=40, which corresponds~to R$\sim$24 for 2--10 keV  
fluxes of\, $\sim$10$^{-14}$\cgs. Most  of the HELLAS2XMM sources have X--ray 
flux higher than this figure.
  
Optical spectra of 59 out of the 110 sources with optical counterparts   
brighter than R=24 have been obtained using EFOSC2 at the ESO 3.6m   
telescope, DOLORES at the TNG and FORS1 at the VLT/UT1 during 6   
observing runs performed between August 2001 and March 2004.  A total of   
12 nights at the 3.6m and TNG telescopes and $\sim 40$ hours of VLT   
time have been devoted to this program.   
   
Long slit spectroscopy has been carried out in the 3800-10000 \AA\   
band with resolution between 7 and 13 \AA. Data reduction was   
performed using both the MIDAS (Banse et al. 1983) and   
IRAF\footnote{IRAF is distributed by the National Optical Astronomy   
Observatories, which is operated by the Association of Universities   
for Research in Astronomy, Inc., under cooperative agreement with the   
National Science Fundation.} packages.  Wavelength calibrations were   
carried out by comparison with exposures of He-Ar, He, Ar and Ne   
lamps. The flux calibration of the spectra was obtained using   
observations of spectro-photometric standard stars (Oke, 1990)   
performed within a few hours from the spectroscopy of our sources.

 
 Regardless of the optical spectra, our discriminant for 
supermassive black hole accretion is high X--ray luminosity; we 
classify a source as an AGN if it has an X--ray luminosity 
log\,$L_{2-10 keV}\geq42$; if a source has an X--ray luminosity 
log\,$L_{2-10 keV}\geq44$ we classify it as a QSO candidate.
Following Fiore et al. (2003), objects with permitted emission lines 
broader than 2000 km/s (FWHM)\footnote{Removing the instrumental 
broadening from the line profile, the adopted velocity threshold 
corresponds to an intrinsic FWHM of $\sim1500$ km/s.} are classified 
type 1 AGN  or QSO; objects with permitted emission lines 
narrower than this threshold and  strong (equivalent width EW$>5$ 
\AA) [OIII], MgII, NeV or CIV emission lines are classified type 2 
AGN; objects with  faint (equivalent width EW$<5$ \AA) [OIII] 
and$/$or strong [OII] or H$\alpha$  and no high ionization 
emission lines are classified Emission-Line Galaxies (ELGs). 
In one case the presence of broad emission lines in the optical 
spectrum cannot be excluded due to the insufficient quality of the 
spectrum and the classification of the object is therefore uncertain 
(see notes in Table 1).  Objects without strong emission lines 
but with stellar absorption lines and a red continuum are classified 
as Early Type Galaxies (ETGs, or XBONGs as they are X--ray bright 
sources with optical spectra typical of early type galaxies).  Table 
1 gives for each source the classification of the optical spectrum, 
the redshift and the X-ray luminosity.

\begin{figure*}   
\centering    
\begin{tabular}{cc}   
\includegraphics[width=8cm]{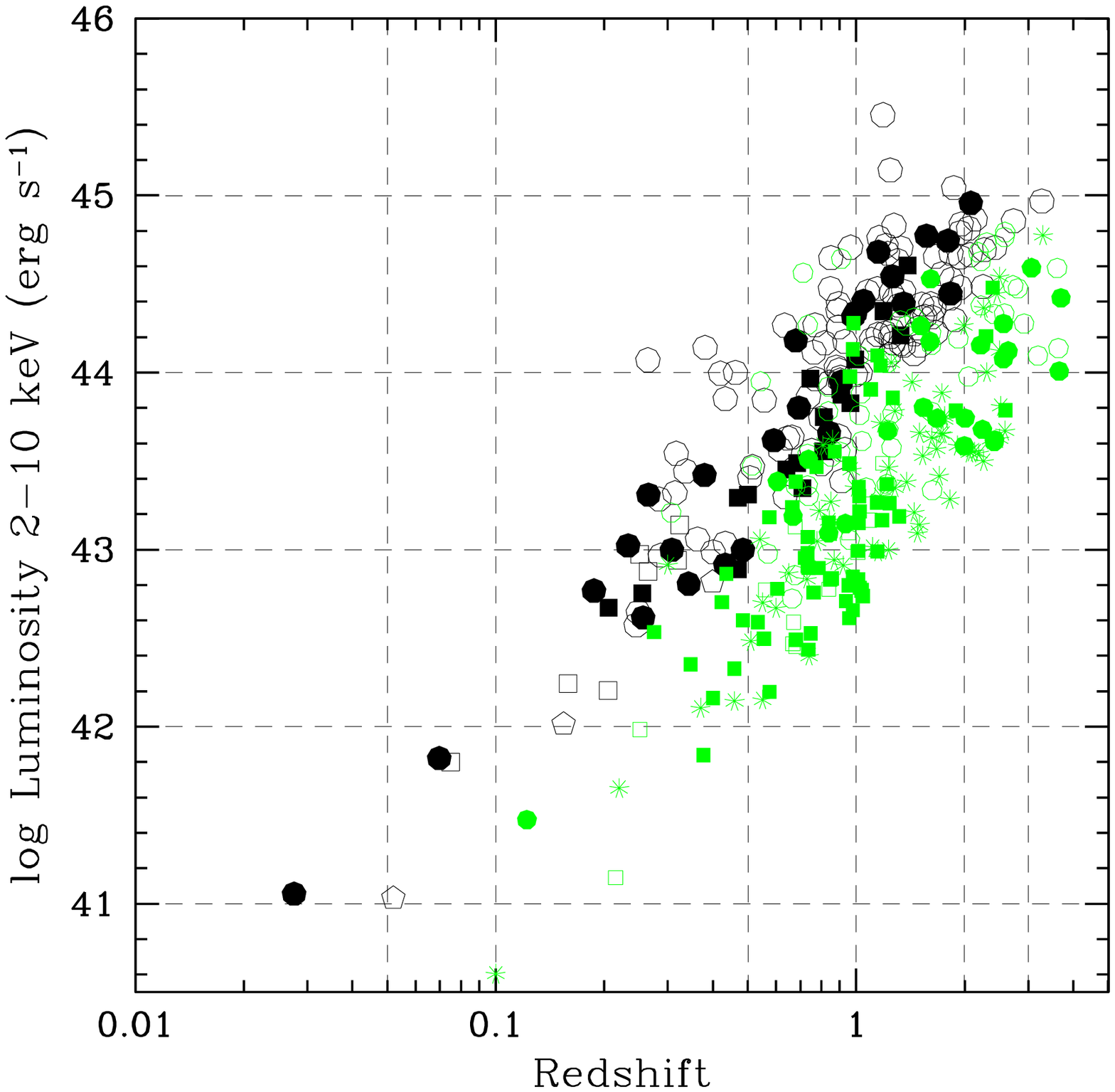}   
\includegraphics[width=8cm]{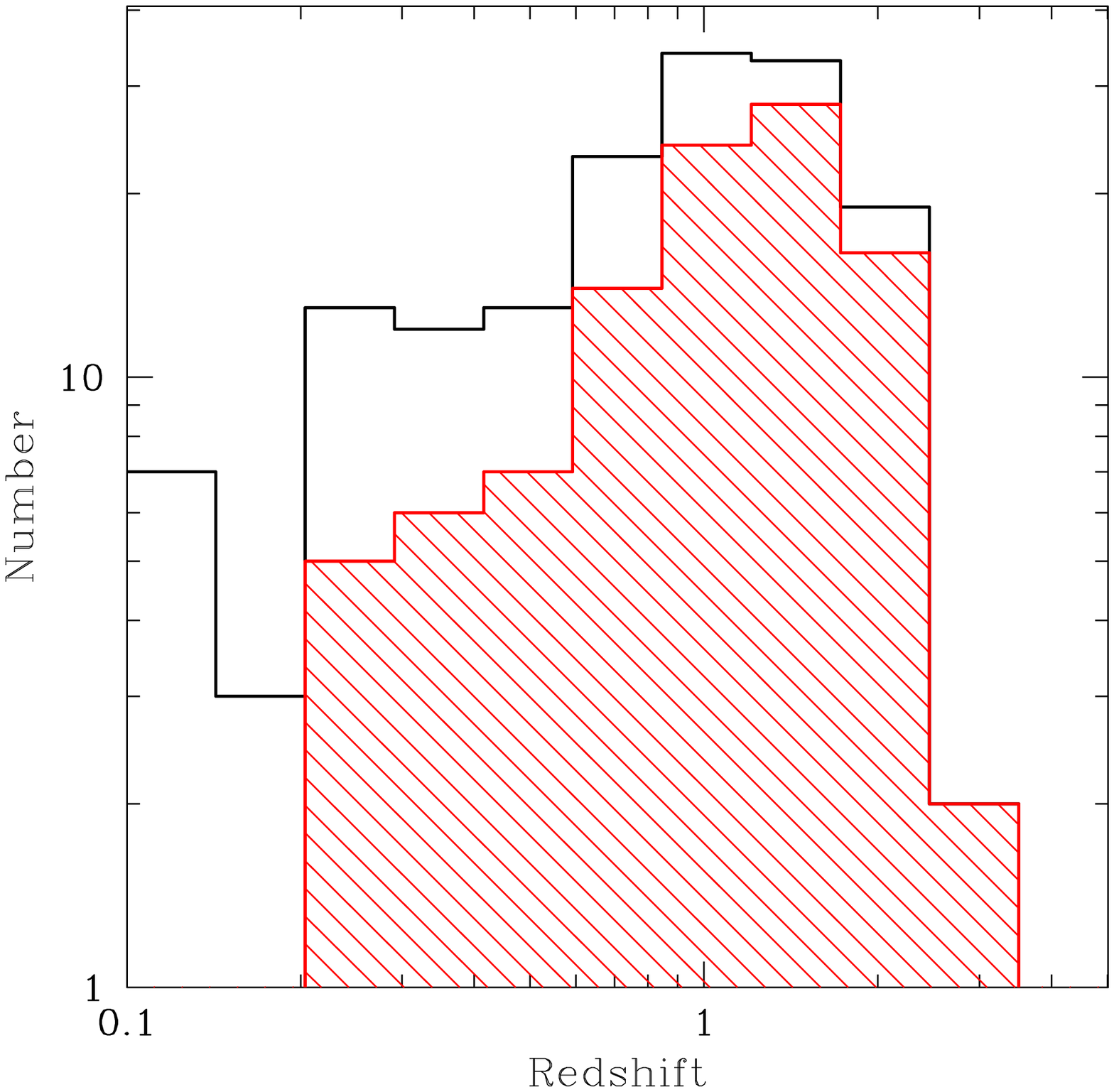}   
\end{tabular}   
\begin{tabular}{cc}   
\includegraphics[width=8cm,height=8cm]{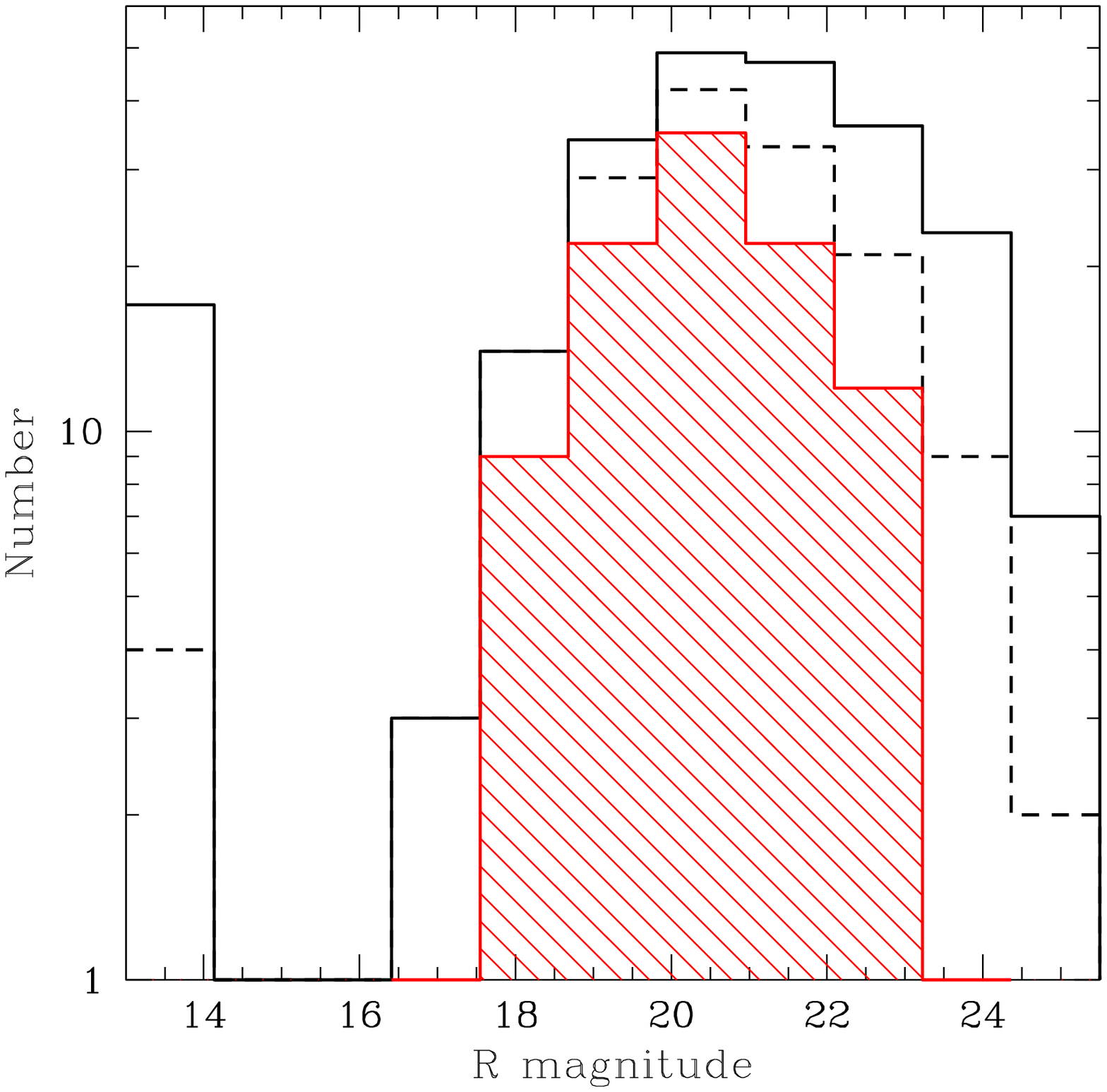}   
\includegraphics[width=8cm,height=8cm]{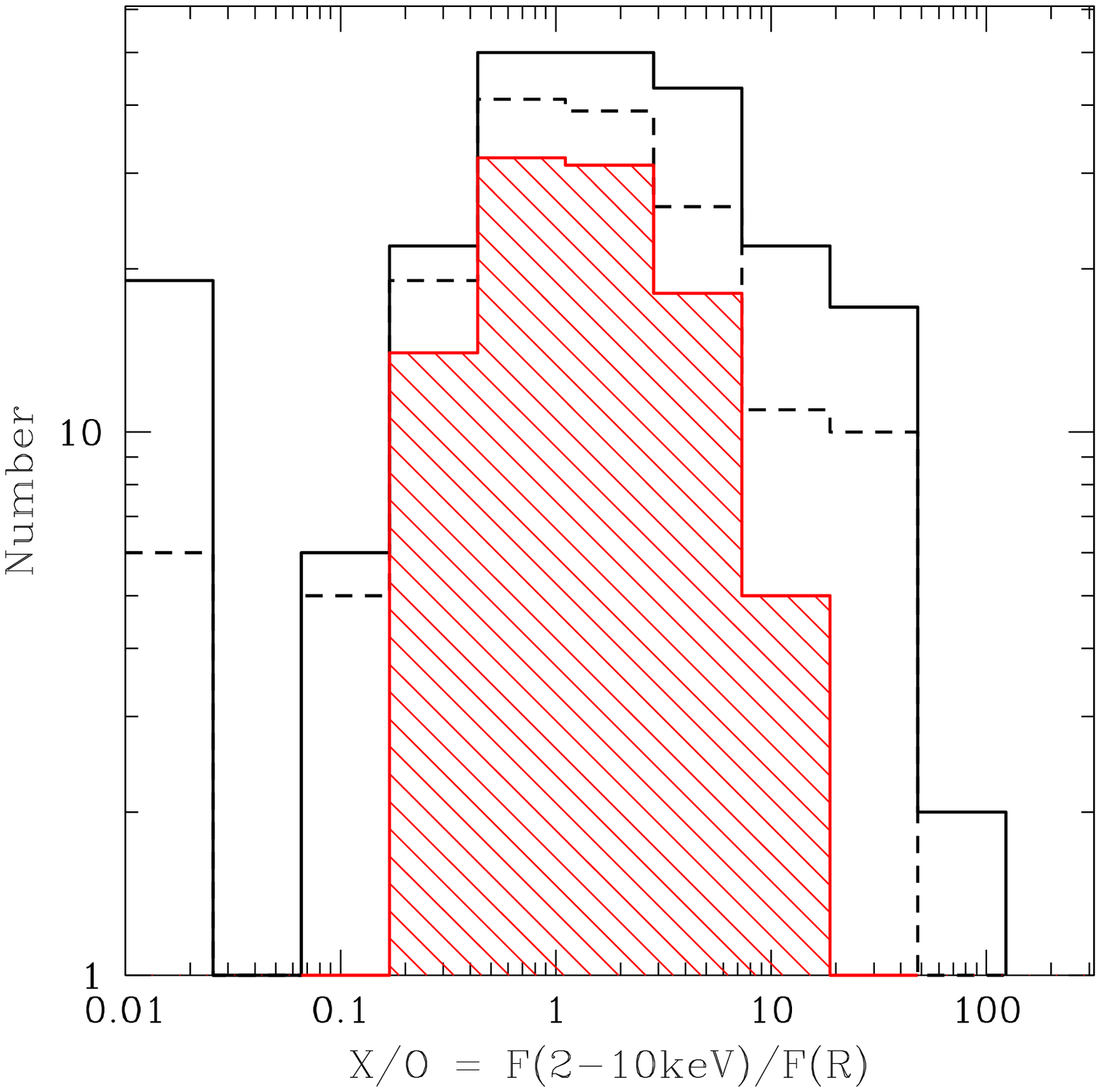}   
\end{tabular}   
\caption{ a) Upper left panel: the 2-10 keV luminosity as a function 
of the redshift for the sources in the full HELLAS2XMM sample (black 
points) compared with the luminosity as a function of the redshift for 
CDFS+CDFN samples (smaller green points;  data from Giacconi et 
al. 2002, Alexander et al. 2003, Barger et al. 2003, Szokoly et 
al. 2004). 
Stars are CDFS and CDFN sources with photometric redshifts.  b) upper 
right panel: the redshift distribution of the 159 sources with 
spectroscopic redshifts of the full HELLAS2XMM sample.  The hatched 
histogram refer to type 1 AGN only and the continuous histogram to the 
source population with spectroscopic z. c) lower left panel: the R 
magnitude distribution; d) lower right panel: the X-ray to optical 
flux ratio X/O distribution.\newline In the c) and d) panels the 
hatched histograms refer to type 1 AGN only, the dashed histograms to 
the source population with spectroscopic z, the continuous histograms 
to the whole source population.} 
\label{zlx}   
\end{figure*}   
   
The source breakdown includes: 41 broad-line AGNs (QSOs and Seyfert 1 
galaxies, log$L_{2-10 keV}>44$ and log$L_{2-10 keV}<44$, respectively); 9 
narrow-line AGNs (type 2 QSOs and Seyfert 2 galaxies); 5 emission-line 
galaxies, all with log$L_{2-10 keV}>42.7$ and therefore all probably 
hosting an AGN; 2 early-type galaxies, both with log$L_{2-10 
keV}>42.0$ and, therefore, XBONG candidates, probably hosting an AGN; 
1 group of galaxies; 1 star.  In summary, 57 of the 59 sources with 
optical spectroscopy are associated with AGN emission, the majority 
(69\%) are type 1 AGNs. This brings the number of confirmed AGNs in the 
full HELLAS2XMM sample to 155 out of 159 with a spectroscopic redshift. 
  
The four panels of figure \ref{zlx} show the redshift-luminosity 
plane, and the redshift, R magnitude, X--ray-to-optical flux ratio 
histograms for the full HELLAS2XMM sample.  The observed fraction of 
type 1 AGNs increases strongly with redshift, corresponding to 
$\sim90\%$ of the {\it whole} source population with a spectroscopic 
redshift higher than 1.5.  Conversely, the fraction of type 1 AGNs 
decreases strongly with increasing X--ray-to-optical flux ratio and 
with the magnitude of the optical counterpart of the X-ray source, 
with only one of the eleven sources with R$\gs$23.3 and a 
spectroscopic redshift identified as a type 1 AGN ( see Fiore et 
al. 2003, La Franca et al. 2005, Eckart et al. 2006, Silverman et 
al. 2005, Treister et al. 2005, Steffen et al. 2004 for similar 
results and more detailed discussions).   This suggests that the 
lack of high--z type 2 AGNs in the sample of spectroscopically 
identified HELLAS2XMM sources (Fig.2, upper right panel) can be 
probably due to the incompleteness of the sample: at z$\gs$1.5 the 
R--band magnitude of the optical counterparts of type 2 AGN is often 
beyond the spectroscopic limit of our sample (R$\sim24$). Furthermore, 
even for objects brighter than the spectroscopic limit, the optical 
nucleus is so dim that the redshift determination is complex for the 
lack of observable lines, the so called spectroscopic redshift desert 
at z=1.5--2. 
 

\subsection{Optically obscured AGN}   
   
\begin{figure*}   
\centering    
\begin{tabular}{cc}   
\includegraphics[width=8cm,height=8cm]{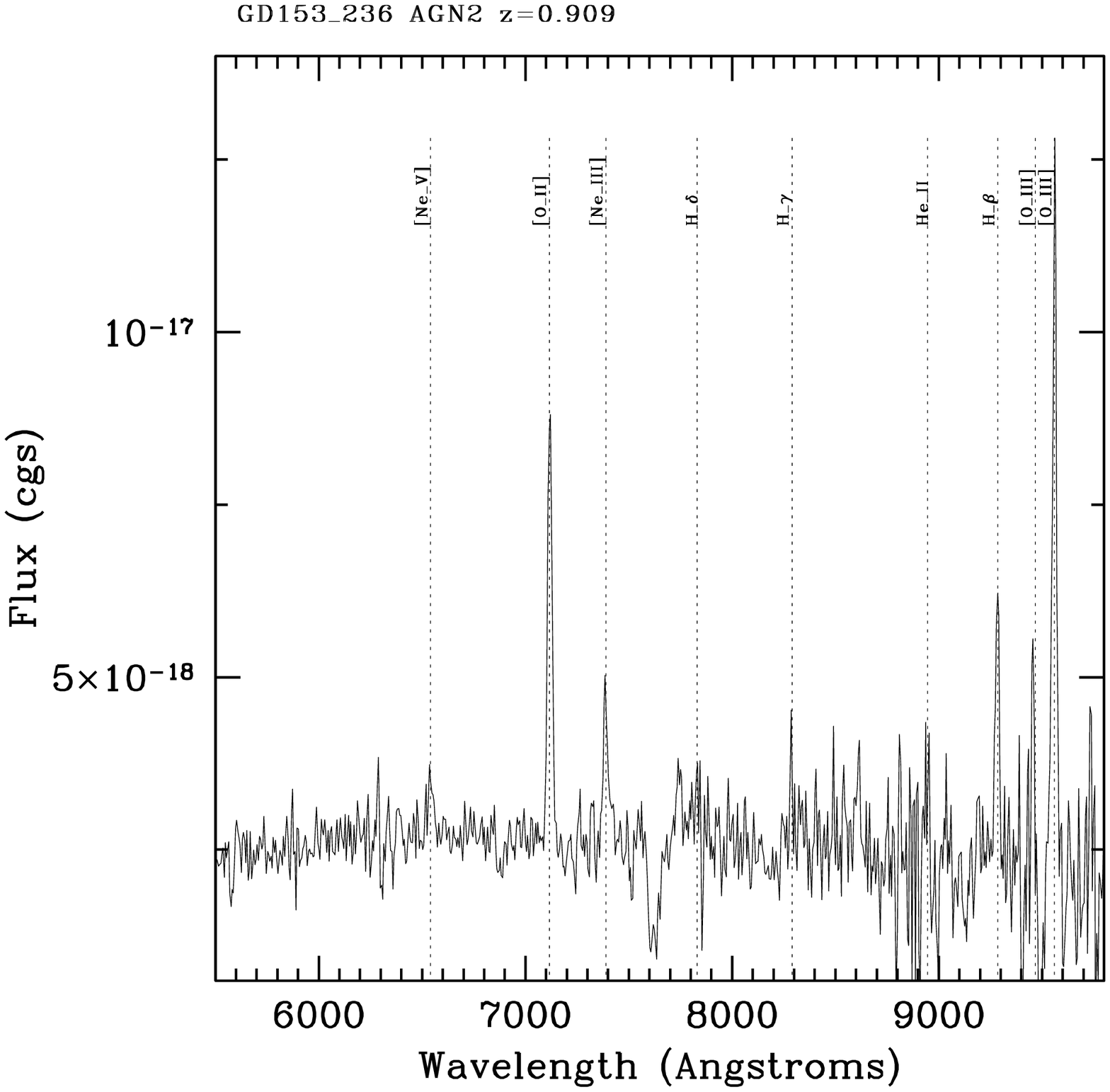}   
\includegraphics[width=8cm,height=8cm]{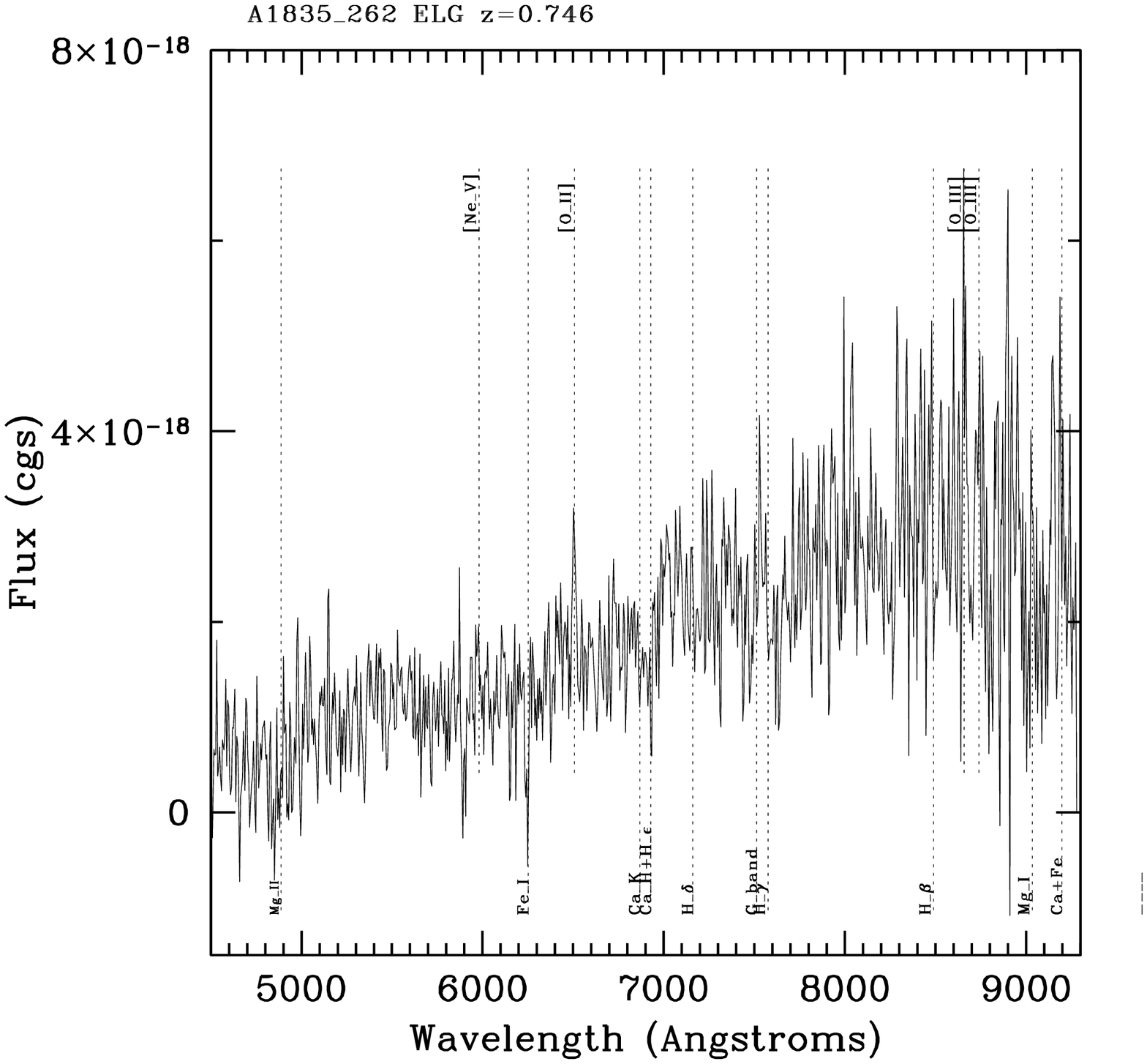}   
\end{tabular}   
\caption{The optical spectra of  GD153\_236 
(H2XMMJ125654.1+215318) and A1835\_140 (H2XMMJ140130.8+024532).  Both 
spectra are classified as narrow emission--line objects based on the 
$H_{\beta}$ FWHM ($<1000$ km s$^{-1}$, GD153\_236), narrow [OII] and 
[OIII] emission, absence of broad permitted lines (A1835\_140), and 
high 2-10 keV luminosity ($\sim10^{44}$ erg s$^{-1}$ for both 
sources).} 
\label{qso2}   
\end{figure*}

 Following Fiore et al. (2003), we limit ourselves to consider two 
broad AGN categories: optically unobscured AGN, i.e. type 1, broad 
emission line AGN, and optically obscured AGN, i.e. non type 1 AGN, in 
which the nuclear optical emission is totally or partly reduced by 
dust and gas in the nuclear region and/or in the host galaxy.  The 
fraction of optically obscured AGN with respect to the total number of 
identified X--ray sources is approximately  27\% (16 out of 59 
X--ray sources, see also the source breakdown).   Two examples of 
such sources are reported in figure \ref{qso2}.  The $H_{\beta}$ FWHM 
of GD153\_236  (H2XMMJ125654.1+215318; z=0.909, AGN2) is $<1000$ 
km s$^{-1}$.  Faint 
[OII] and [OIII] emission is present in the spectrum of A1835\_262 
 (H2XMMJ140130.8+024532; z=0.746, ELG). The latter source is 
obscured also in the X-ray band, having a 0.5-2 keV flux less than 20 
times the observed 2-10 keV flux;  to be more quantitative, the 
X--ray spectral fit yields a rest frame N$_H$(z)$\simeq$ 
$7.9^{+3.3}_{-2.2} \times 10^{22}$\,~cm$^{-2}$, fixing the spectral 
energy index $\alpha_E$ (F(E)$\propto$ E$^{-\alpha_E}$) to 0.9. 
GD153\_236 is less extreme in this regard,  the X--ray spectral 
fit yields a rest frame N$_H$(z)~=~$0.3^{+0.4}_{-0.3} \times 
10^{22}$\,~cm$^{-2}$, fixing the spectral energy index $\alpha_E$ to 
0.9. 
The 2--10 keV  intrinsic luminosity is $1.10 \times 10^{44}$ erg 
s$^{-1}$ for A1835\_262 and $1.59 \times 10^{44}$ erg s$^{-1}$ for 
GD153\_236, making them  obscured QSO candidates. 
 
 We are interested in finding a statistical method to select 
highly obscured QSOs candidate among the still unidentified sample 
sources in order to estimate their surface density. To this purpose, 
in the following, we will make use of observed quantities, such as 
the observer--frame absorbing column N$_H$ and the X/O ratio. We use 
the results of our spectroscopic identifications to calibrate and 
validate this statistical method.
  
\begin{figure*}   
\centering   
\begin{tabular}{cc}   
\includegraphics[width=8cm]{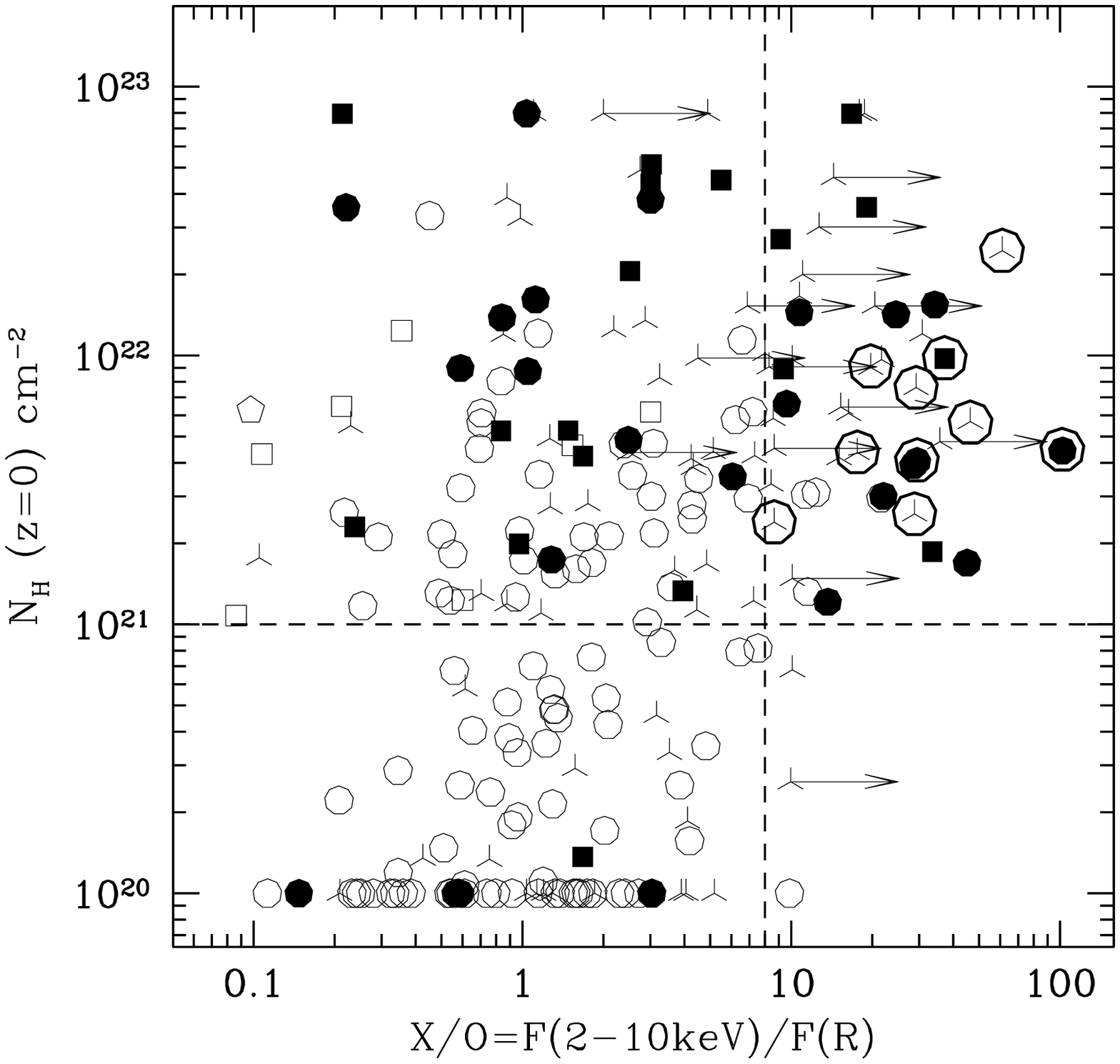}   
\includegraphics[width=8cm]{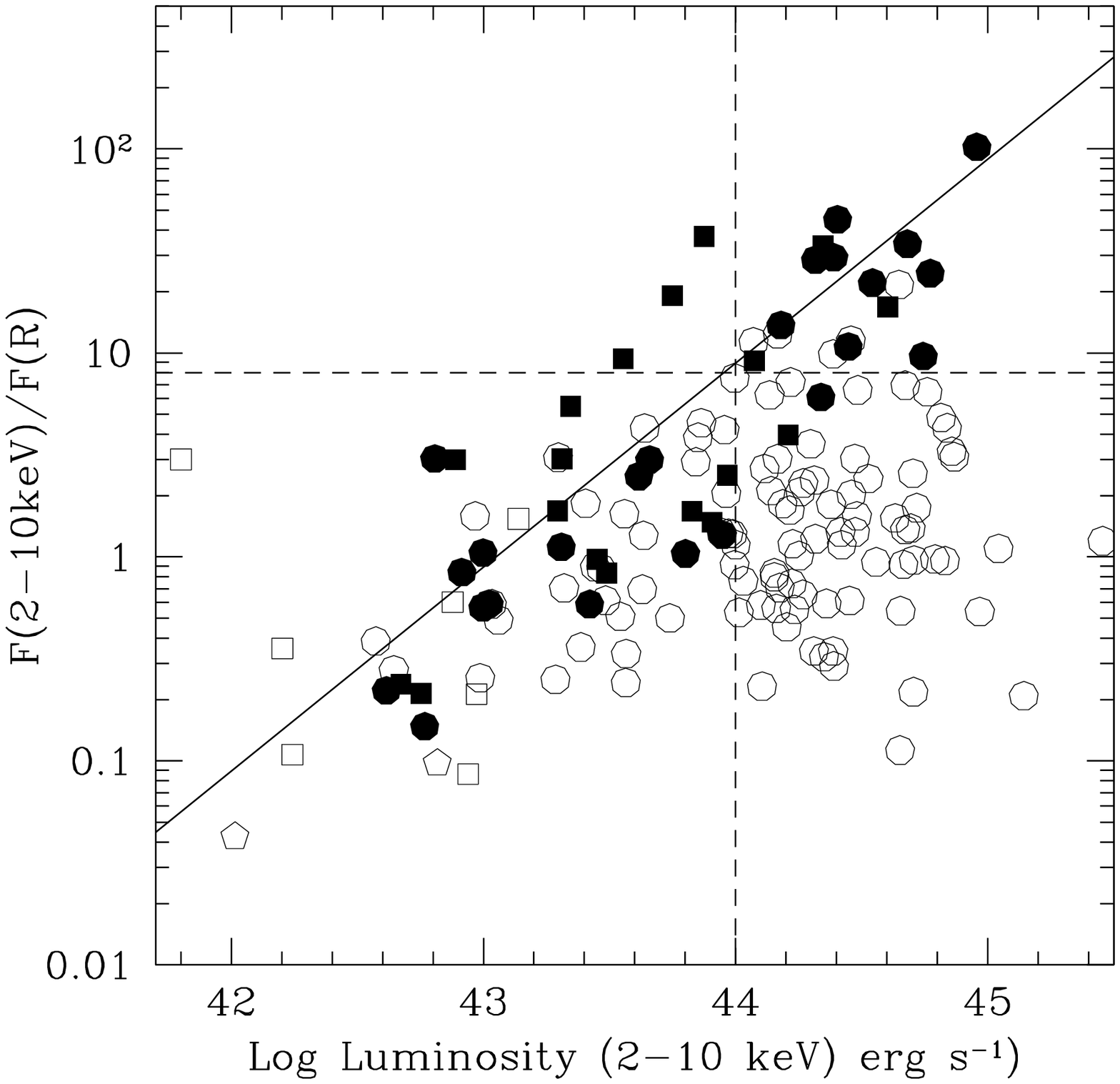}   
\end{tabular}{cc}   
\caption{  The observed-frame absorbing column N$_H$(z=0) as a 
function of the X-ray (2-10 keV) to optical (R$-$band) flux ratio X/O 
(left panel) and the X/O ratio as a function of the 2-10 keV 
luminosity (right panel) for the HELLAS2XMM full sample. Open circles 
= broad-line AGN; filled circles = narrow-line AGN; filled squares = 
emission-line galaxies; open squares = early-type galaxies; pentagons 
= groups/clusters of galaxies; skeleton triangles = unidentified 
objects. The 10 encircled symbols in the left panel have extreme R--K 
colors (R-K$>5$, see Mignoli et al. 2004). The diagonal line, in the 
right panel, is the best linear regression between log (X/O) and log 
L(2-10 keV) (from Fiore et al. 2003).  The dashed lines represent the 
loci of constant $N_H = 10^{21}$ cm$^{-2}$, X/O=8 and log L(2-10 
keV)=44.} 
\label{xonho}   
\end{figure*}

The observed-frame absorbing column N$_H$(z=0) (computed from the 
hardness ratios following Fiore et al. 2003) as a function of X/O for 
the full HELLAS2XMM sample is reported in figure \ref{xonho} (left 
panel).  We note that the part of the diagram at X/O$>$8 and 
log\,N$_H$(z=0) $>21$ is populated mainly by narrow-line objects. 
More quantitatively, 16 out of the 20 objects with optical 
spectroscopy are, according to the definition of Fiore et al. (2003), 
optically obscured AGNs (i.e., $80\pm20\%$). 
 
\begin{figure*}   
\centering   
\begin{tabular}{cc}  
\includegraphics[width=8cm]{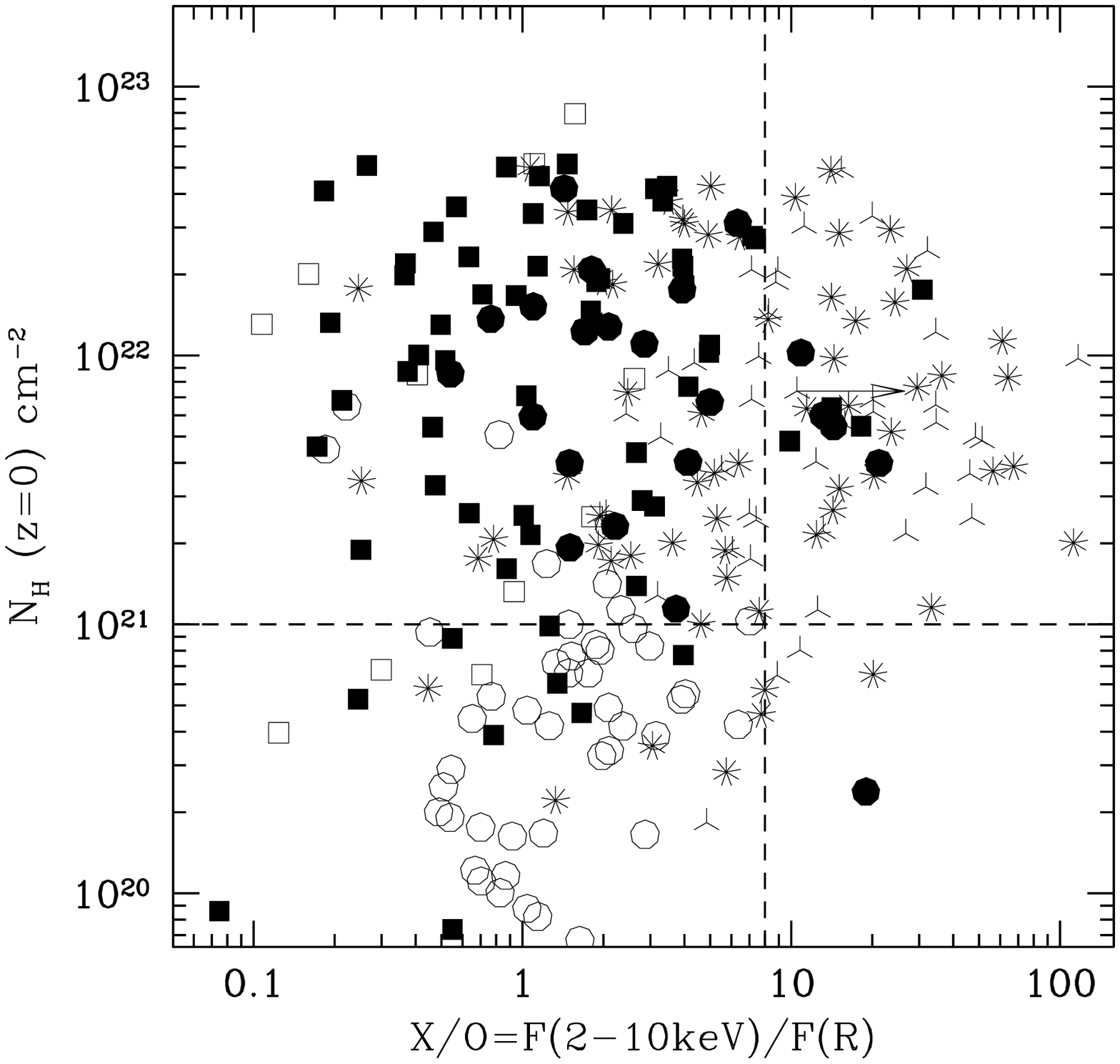}   
\includegraphics[width=8cm]{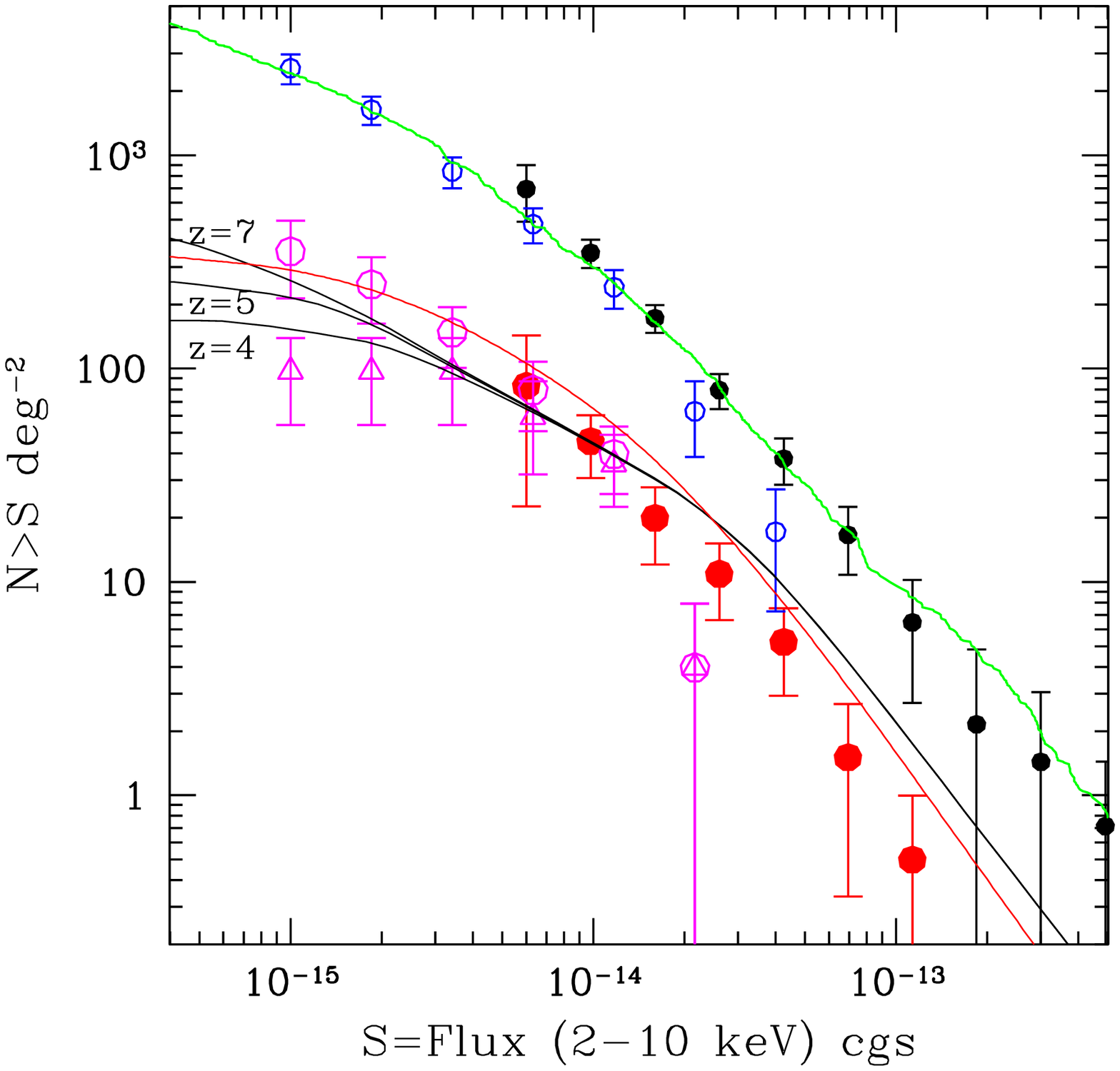}   
\end{tabular}  
\caption{ a) Left panel:  the observed-frame absorbing column 
N$_H$(z=0) as a function of the X-ray (2-10 keV) to optical (R$-$band) 
flux ratio X/O for the CDFS and CDFN samples.  Open circles = 
broad-line AGN; filled circles = narrow-line AGN; filled squares = 
emission-line galaxies; open squares = early-type galaxies; pentagons 
= groups/clusters of galaxies; skeleton triangles = unidentified 
objects. The stars are sources with a photometric redshift. The dashed 
lines represent the loci of constant $N_H =10^{21}$ cm$^{-2}$ and 
X/O=8. 
b) right panel: the X-ray (2-10 keV) 
integral number counts from the HELLAS2XMM (filled symbols) and 
CDF+CDFN samples (open symbols). The upper points are the number 
counts of the full samples, the solid curve is from the Moretti et 
al. (2003) compilation.  The lower points refer to high-luminosity 
(logL(2-10 keV)$>44$), highly obscured sources only.  Open triangles 
include only sources with optical counterparts brighter than 
R=25.5. Black continuous lines represent the expected number counts 
from the Comastri et al. (2001) pure luminosity evolution XRB 
synthesis models (z=4, 5, 7 are the maximum redshift over which the 
integration of the XLF is performed); the red line represents the 
expected number counts from the La Franca et al. (2005) luminosity 
dependent density evolution model.  } 
\label{xolx}   
\end{figure*}

Since the observed log\,N$_H$=21 would correspond to a rest-frame 
column density of log\,N$_H$ = 21.5 at $z$=1, most of these AGNs, 
having redshifts z$\gs$1, are also X--ray obscured (see also Mainieri 
et al. 2002).  The right panel of figure \ref{xonho} shows X/O as a 
function of the 2-10 keV luminosity for the HELLAS2XMM full source 
sample. This figure clearly shows that most optically obscured AGN 
with X/O$>$8 have high-luminosities (log\,~L~(2-~10~keV)~$>44$), 
highlighting the efficiency of a selection based on the observed X/O 
ratio to find highly obscured QSO candidates.\\  The number of 
spectroscopically identified sources in the HELLAS2XMM sample with 
log\,N$_H$(0)$>$21, X/O$>$8 and 
log\,L(2-10 keV)$>44$ is 17. 13 of these are optically obscured AGN, 
i.e. a fraction of 0.76$\pm0.25$.\\ The diagram in the left panel 
of figure \ref{xolx} shows the N$_H$(0) as a function of X/O ratio for 
the CDFS and CDFN sources ( Giacconi et al. 2002, Alexander et al. 
2003, Barger et al. 2003, Szokoly et al. 2004) selected by Fiore 
(2004) by having 2-10 keV fluxes larger than $10^{-15}$ \cgs and off 
axis angles lower than 10 arcmin.  Unfortunately, most of the sources 
from these samples with X/O$>$8 and log\,N$_H$(z=0)$>21$ do not have a 
spectroscopic redshift\footnote{As regard the optical classification 
of the CDFS and CDFN identified sources we applied, where possible, 
our classification criteria. The X--ray fluxes are taken from 
Alexander et al. (2003). The catalog reports 2--8 keV fluxes converted 
to 2-10 keV fluxes using a power--law model with spectral energy index 
$\alpha_E$=0.8.}: in the following analysis we will assume that the 
fraction of optically obscured AGN in these faint sources is similiar 
to the fraction in the HELLAS2XMM sample, i.e. a fraction of 
0.76$^{+}_{-}0.25$.  Assuming the fraction of optically obscured 
QSOs, it is possible to compute their integral number counts 
relations which are compared with the logN--logS of the entire samples 
in figure \ref{xolx}b).  Number counts and relative sky coverages are 
also tabulated in Table 2.

The line of reasoning followed to compute the fraction of optically 
obscured QSOs implies that the Fiore et al. (2003) relationship 
between X/O and log\,L(2--10 keV) may be directly applicable to the 
CDFS and CDFN sources with the faintest optical counterparts, assuming 
that also these sources are very likely high-luminosity objects, 
including them among the optically obscured QSO population.  However, 
Bauer et al. (2004) and Barger et al. (2005) pointed out that the 
Fiore et al. (2003) relationship could not be valid at faint (R $>$ 
25.5) optical magnitudes. 
For this reason, we re-computed the optically obscured QSO number counts by 
excluding all X-ray sources with optical counterparts fainter than 
R=25.5 (open triangles in figure \ref{xolx}b). This is probably a 
lower limit to the "true" type 2 QSO surface density.  In summary, we 
find a highly obscured QSO density of 45$\pm15$ and 100--350 deg$^{-2}$ at flux limits 
of $10^{-14}$ and $10^{-15}$ \cgs, respectively (see figure 
\ref{xolx}b). The fraction of obscured QSOs to the total number of 
X--ray sources selected in the 2--10 keV band is therefore $\sim 13\%$ 
and between 4--14\% at the two flux limits.

\setcounter{table}{1}   
   
\begin{table*}[ht!]   
\caption{\bf Number counts and sky coverage}   
\begin{tabular}{lcccc}   
\hline   
\hline   
F(2-10 keV) & Sky-coverage & Tot. N.C. & QSO2 N.C. & QSO2 N.C.(R$<$25.5) \\   
($10^{-14}$\cgs)       &  (deg$^{2}$)   & (deg$^{-2}$)& (deg$^{-2}$)&  (deg$^{-2}$)  \\   
\hline   
\multicolumn{5}{c}{HELLAS2XMM}\\   
\hline   
11.3 & 1.39  & 5.7$\pm$2.8 & 0.5$\pm$0.5 & \\   
6.93 & 1.37  &  16$\pm$5.1 & 1.6$\pm$1.4 & \\   
4.25 & 1.25  &  38$\pm$8.7 & 5.7$\pm$3.1 & \\   
2.60 & 0.98  &  79$\pm$15  &  12.0$\pm$6.1 & \\   
1.60 & 0.5   & 170$\pm$26  &  21.9$\pm$11.3  & \\   
0.98 & 0.11  & 350$\pm$53  &  50.1$\pm$23.2  & \\   
0.60 & 0.015 & 700$\pm$210 &  91.1$\pm$73.0  &   \\   
\hline   
\multicolumn{5}{c}{CDFS+CDFN}\\   
\hline   
4.00 & 0.174 &    17$\pm$9.9 &       -       &       -       \\   
2.16 & 0.174 &    63$\pm$24  &   4.3$\pm$4.6 &    4.3$\pm$4.6 \\   
1.17 & 0.174 &   240$\pm$50  &  43$\pm$21  &    39$\pm$19  \\   
0.63 & 0.174 &   475$\pm$89  &   87$\pm$43  &    65$\pm$37  \\   
0.34 & 0.14  &   840$\pm$140 &   162$\pm$74  &   106$\pm$58  \\   
0.18 & 0.07  &  1640$\pm$250 &   272$\pm$130 &   106$\pm$58  \\   
0.10 & 0.06  &  2560$\pm$410 &   388$\pm$200 &   106$\pm$58  \\   
\hline   
\end{tabular}   
   
\end{table*}   
   
\subsection{X--ray Bright Optical Normal Galaxies} 
 
\begin{figure*}   
\centering    
\begin{tabular}{cc}   
\includegraphics[width=8cm]{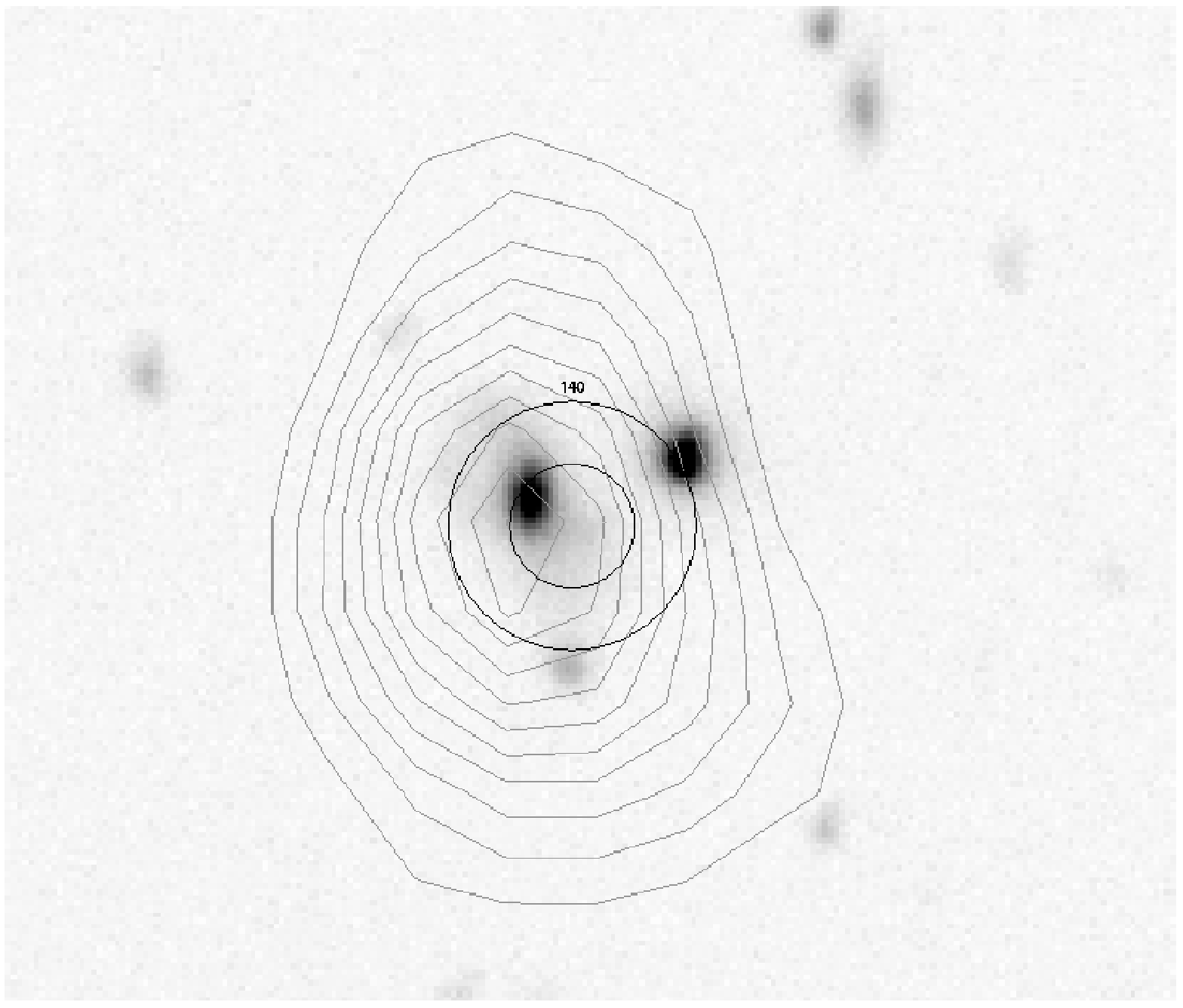}   
\includegraphics[width=8cm]{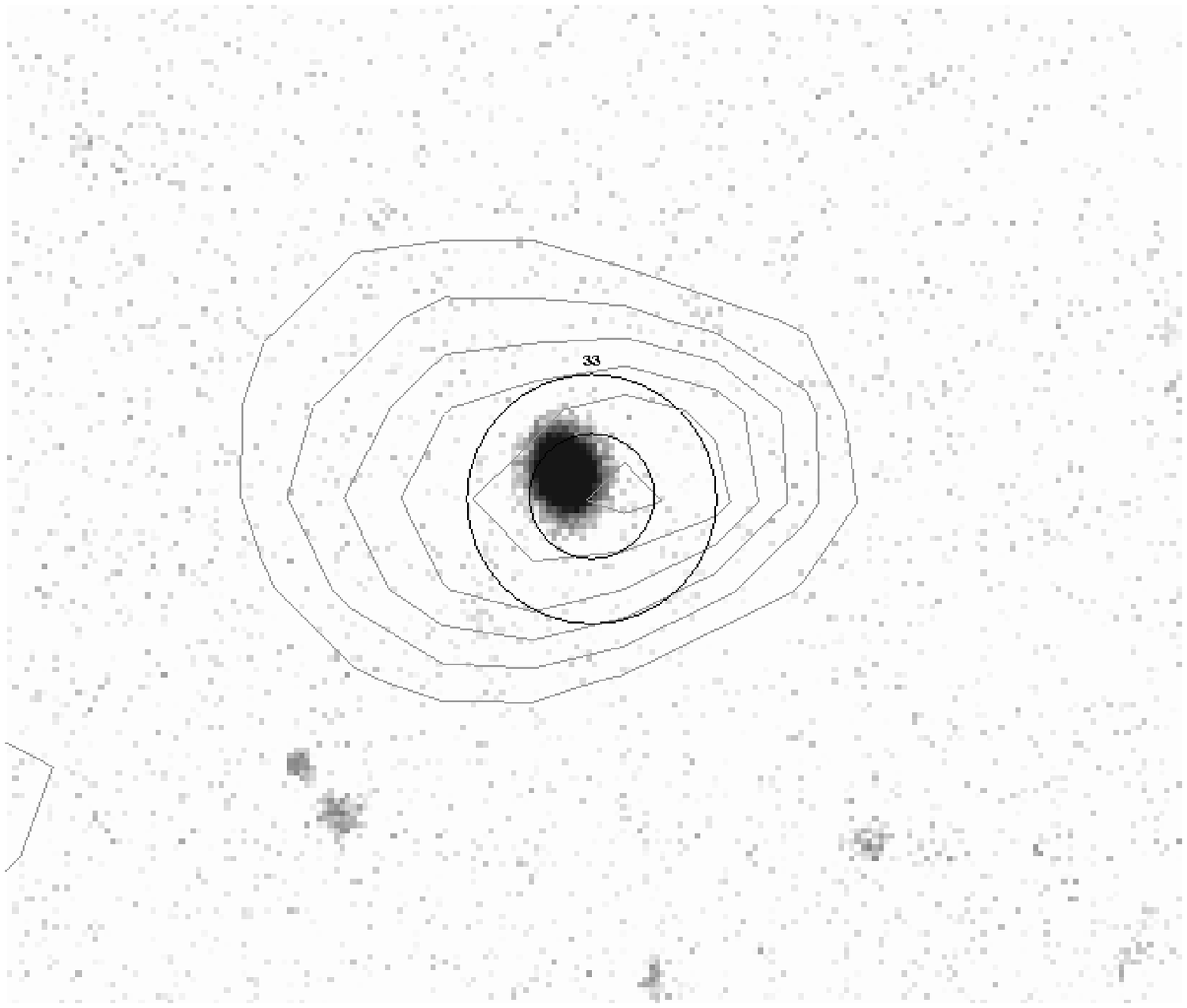}   
\end{tabular}   
\begin{tabular}{cc}   
\includegraphics[width=8cm,height=7cm]{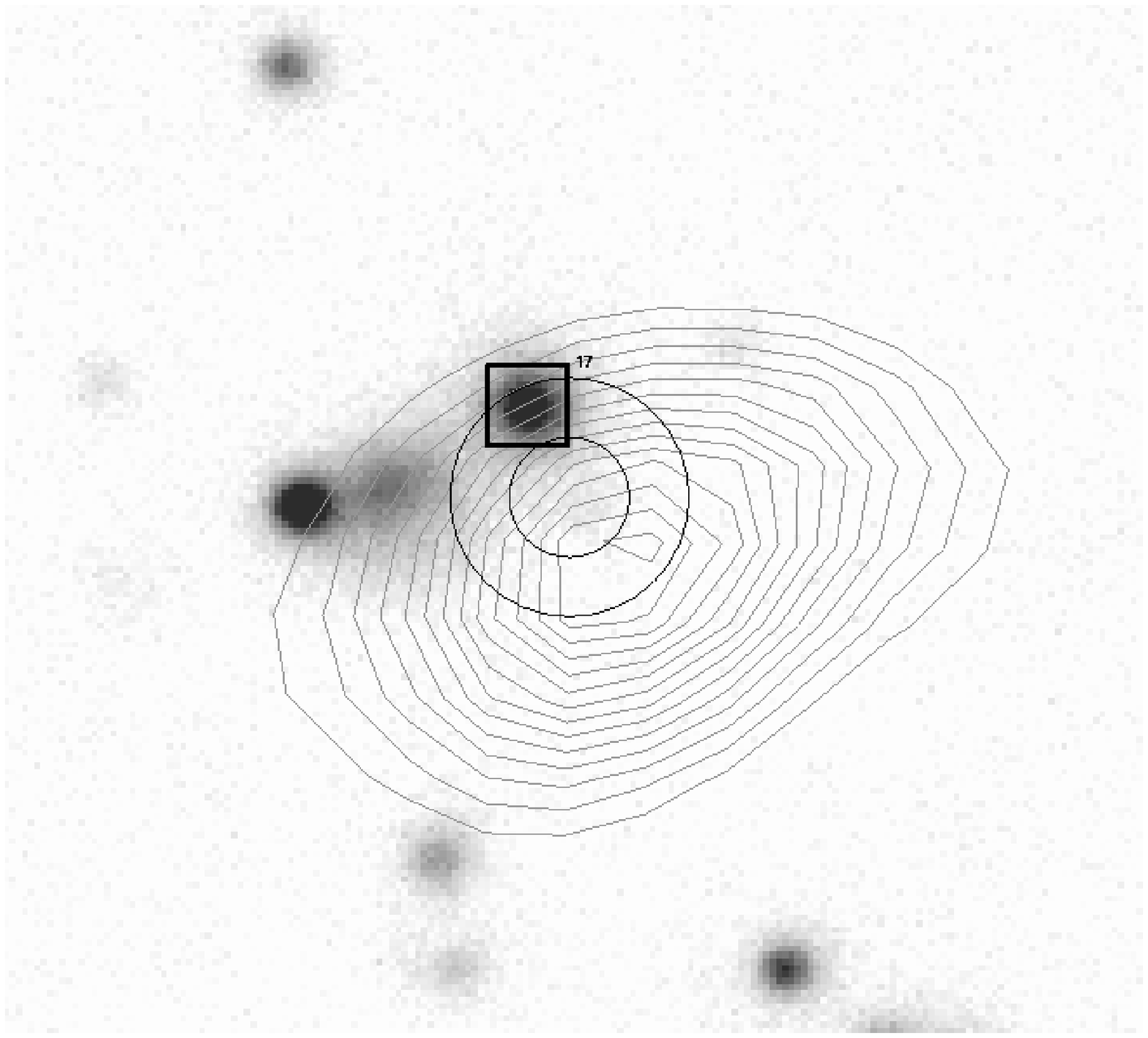}   
\includegraphics[width=8cm,height=7cm]{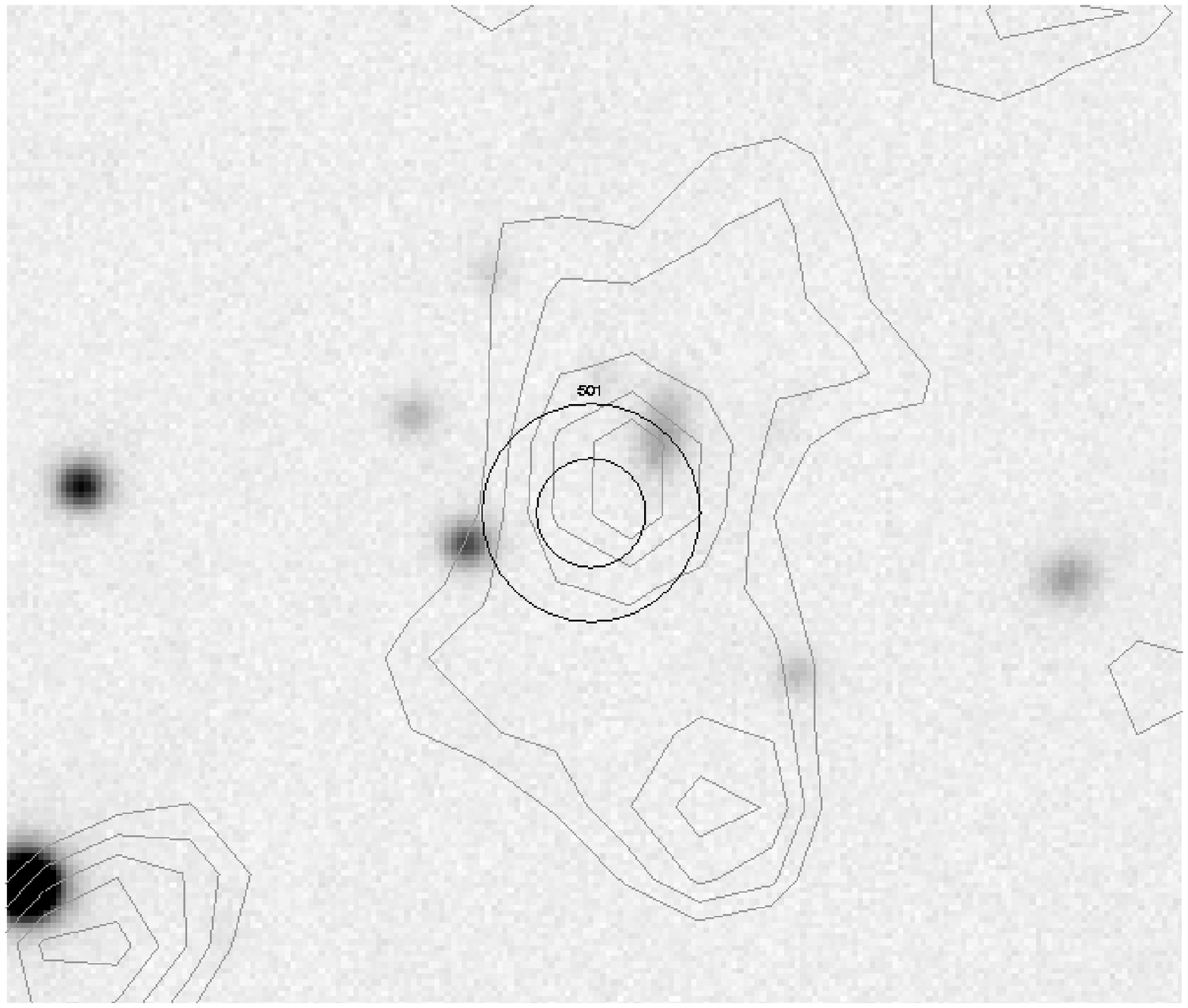}   
\end{tabular}   
\caption{  Upper panels: the R-band images of A1835\_140 
(H2XMMJ140144.9+025332, left) and A1835\_33 (H2XMMJ140057.3+023942, 
right): superimposed are the X--ray position error boxes (3'' and 6'' 
black circles) and the X--ray contours (grey).  The X-ray contours are 
well centered on the sources identified by us as the optical 
counterpart of the X-ray sources.  Bottom panels: the R-band images of 
PKS0312\_17 (H2XMMJ031124.8-770139, left) and PKS0312\_501 
(H2XMMJ030952.2-764927, right): the optical counterpart identified for 
PKS0312\_17 is coincident with a relatively bright radio source (the 
square refers to the radio centroid, 4'' width, see Brusa et 
al. 2003).  The X-ray centroid of PKS0312\_501 is slightly shifted 
from the peak of the X-ray contours, which are coincident with the 
optical countepart. We therefore consider also this indentification as 
reliable. } 
\label{xbong}   
\end{figure*}

\begin{figure*}   
\centering    
\begin{tabular}{cc}   
\includegraphics[width=8cm,height=8cm]{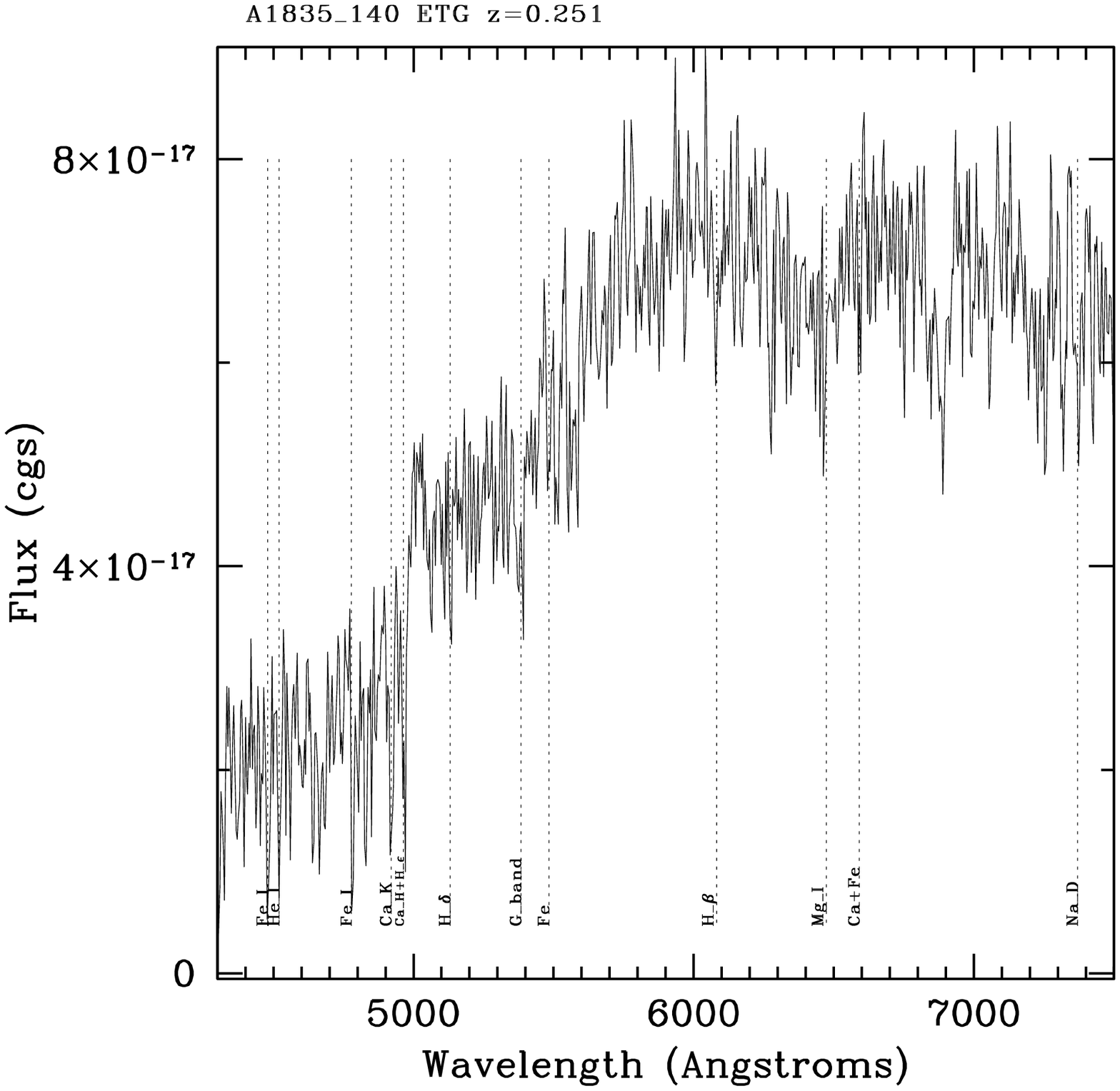}   
\includegraphics[width=8cm,height=8cm]{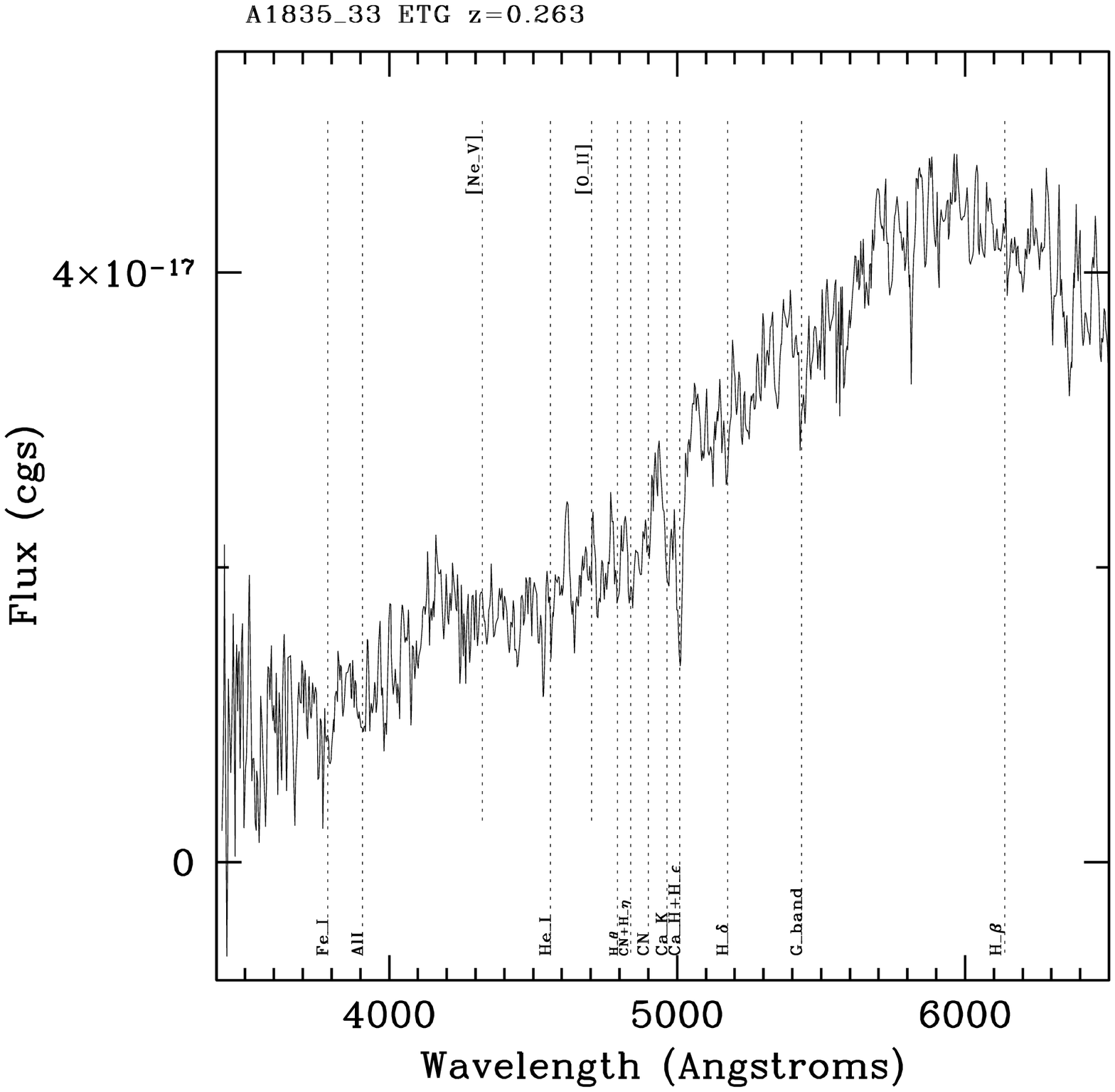}   
\end{tabular}    
\caption{The optical spectra of A1835\_140 (H2XMMJ140144.9+025332, z=0.251) and A1835\_33 (H2XMMJ140057.3+023942, 
z=0.263).  Note the strong Balmer breaks, red continua and absence of 
strong [OII] and [OIII] emission lines.} 
\label{xbong0}   
\end{figure*}

 In figure \ref{xbong} we show the R--band images of the XBONG 
candidates with the largest X--ray to optical position difference. 
Superimposed are the X--ray position error boxes (3'' and 6'' black 
circles) and the X--ray contours (grey).\\ Close to the X-ray position 
of A1835\_140 there are 2 optically bright sources, one within 3'' 
from the X--ray centroid and the other at 6.6''; both sources are 
clearly extended and their morphology is consistent with a spheroid. 
No optical point--like nucleus is evident in both sources (see, also, 
Civano et al. 2006, in preparation).\\ The X-ray contours are well 
centered on the source we identified as the optical counterpart of the 
X-ray source.\\ The R-band images of PKS0312\_17 
(H2XMMJ031124.8-770139) and PKS0312\_501 (H2XMMJ030952.2-764927) are 
reported in the bottom panels; the optical counterpart identified for 
PKS0312\_17 lies at 4.6'' from the nominal X--ray centroid but it is 
coincident with a relatively bright radio source (the square refers to 
the radio centroid, 4'' width, see Brusa et al. 2003 for details). We 
therefore consider this optical source as the most likely counterpart 
of the X--ray source.  As we can see in fig.~\ref{xbong}, the X-ray 
centroid of PKS0312\_501 provided by the detection alghorithm is 
slightly shifted from the peak of the X--ray contours, which is 
coincident with the optical countepart. We therefore consider as 
reliable also this identification.\\ A paper discussing in more detail 
these sources and presenting near--infrared imaging is in preparation 
(Civano et al. 2006).  As an example, figure \ref{xbong0} shows the 
spectra of 2 XBONGs, which we tentatively classify as early-type 
galaxies based on the strong red continuum and the absence of emission 
lines (however, note that these spectra do not cover the H$\alpha$ 
transition). 
 
\subsection{[OIII] emission}   
   
For 59 of the 159 optical spectra of the full HELLAS2XMM sample the 
[OIII]$\lambda5007$ line falls in the observed wavelength range.  We 
measured\footnote{The [OIII]$\lambda5007$ fluxes were measured 
performing a Gaussian fit of the line using the standard IRAF task 
{\it splot}.}  significant [OIII]$\lambda5007$ emission in 49 cases: 
26 broad-line AGNs, 23 narrow-line AGNs (including 8 objects with 
optical spectra classified as emission-line galaxies but with 2-10 keV 
luminosity higher than $10^{42}$ erg s$^{-1}$). 
In the 
remaining 10 cases, 7 candidate XBONGs  and 3 galaxies belonging to 
groups or small clusters of galaxies, we computed the 3$\sigma$ upper 
limits (see Table 3).  
   
\begin{figure*}   
\centering    
\begin{tabular}{cc}   
\includegraphics[width=8cm,height=8cm]{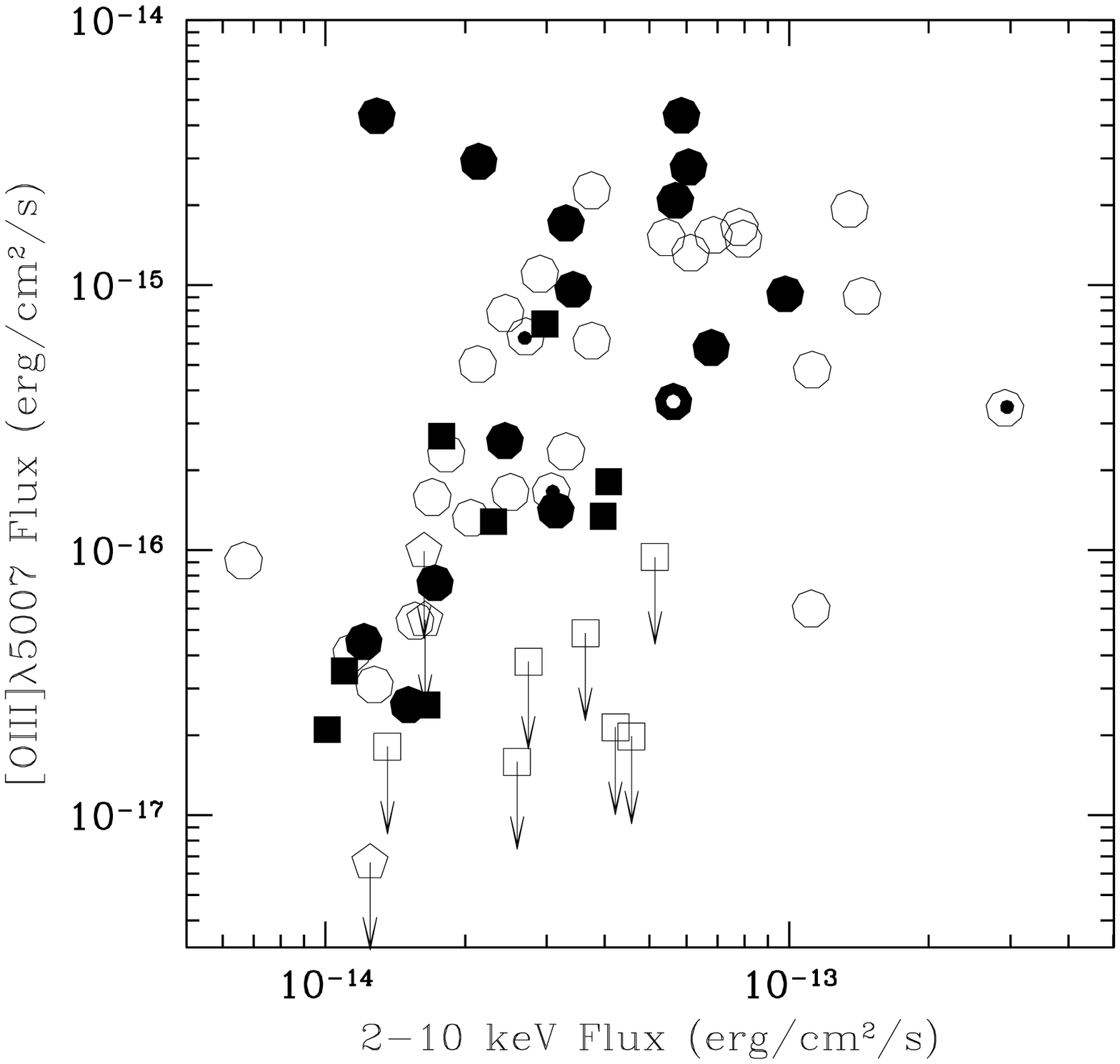}   
\includegraphics[width=8cm,height=8cm]{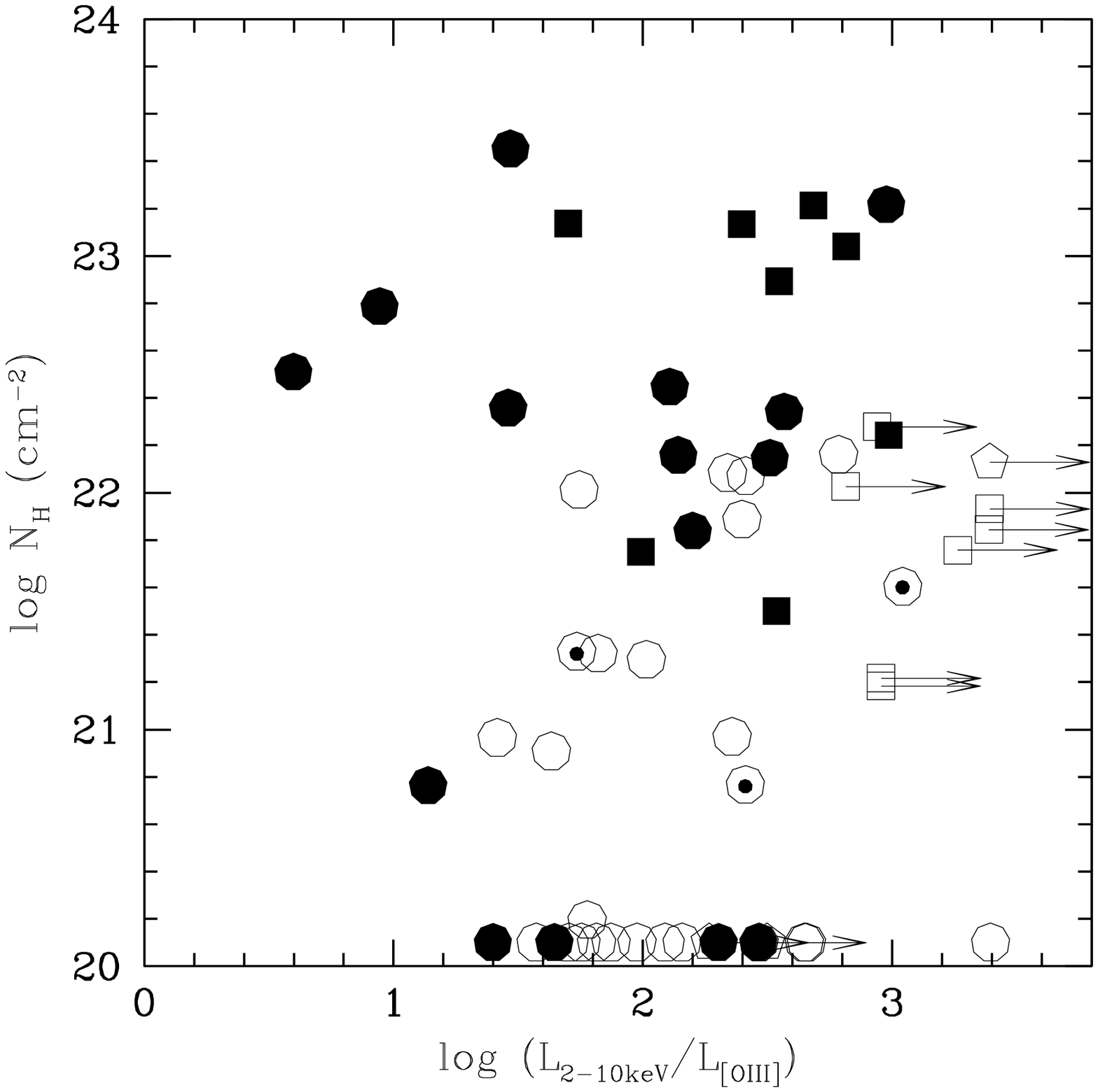}   
\end{tabular}   
\caption{a) Left panel: the [OIII]$\lambda5007$ flux as a function of 
the 2-10 keV flux for 59 sources of the full HELLAS2XMM sample. b) 
Right panel: the rest-frame absorbing column density N$_H$ (computed 
using flux ratios) as a function of the ratio between the 2-10 keV 
luminosity and the [OIII]$\lambda5007$ luminosity.   Open circles 
= broad-line AGN; filled circles = narrow-line AGN; filled squares = 
emission-line galaxies; open squares = early-type galaxies; pentagons 
= groups/clusters of galaxies; skeleton triangles = unidentified 
objects. The four encircled symbols in both panels mark objects for 
which the [OIII]$\lambda5007$ line lies within strong telluric 
absorption features.} 
\label{oiii}   
\end{figure*}

Figure \ref{oiii}a) shows the [OIII]$\lambda5007$ flux as a function 
of the 2-10 keV flux, while figure \ref{oiii}b) shows the rest-frame 
N$_H$ (computed using flux ratios, see Fiore et al. 2003) as a 
function of the ratio between the 2-10 keV and the [OIII]$\lambda5007$ 
luminosities for the 59 objects. Note that the XBONG candidates, 
 in figure \ref{oiii}a), have sistematically lower [OIII] fluxes 
compared with other sources with the same X--ray flux and have 
$L_{2-10keV}/L_{[OIII]}\gs1000$.  The only other HELLAS2XMM AGN
with a X--ray--to--[OIII] luminosity ratio well above this value is
the broad line AGN A2690\_3, a Seyfert 1 galaxy at z=0.433 with a
strong broad MgII and $H_{\beta}$ lines but very faint [OIII] 
emission\footnote{See http://www.bo.astro.it/$\sim$hellas/sample.html}. Other three
HELLAS2XMM AGN have $L_{2-10keV}/L_{[OIII]}\sim1000$. This suggests
that AGNs with small [OIII] emission are probably more common than
what was thought before and the the XBONGS are the tip of the iceberg
of these source population.

\section{Discussion}   
   
The hard X-ray selection and the good photometric and spectroscopic 
coverage of the ten HELLAS2XMM fields allow us to probe AGN activity 
over a wide range of broad band properties.  In this paper we focus 
our attention on two extremes of the AGN activity: on one side, 
sources with high X--ray-to-optical flux ratio, the majority of which 
turned out to be the so far ``elusive'' high-luminosity, highly 
obscured type 2 QSOs; on the other side, sources with relatively low 
X--ray-to-optical flux ratios, moderately luminous ($L_X \simeq 
10^{42-43}$ erg s$^{-1}$) active nuclei in otherwise inactive 
galaxies, the so called XBONGs. 

\subsection{The high X/O flux ratio sources} 
As far as the first topic is concerned, we confirm the finding that 
about 20\% of hard X--ray selected sources have an X--ray-to-optical 
flux ratio which is, on average, one order of magnitude or more higher 
than that of optically selected AGNs.  Our spectroscopic 
identifications  suggest that $\sim$  76\% of the sources 
with X/O$>$8 and log\,N$_H$(z=0)$>21$  are highly obscured QSOs 
at $z$=0.7--2  (see also Table 1). 
This implies a number density of optically obscured QSOs of $\sim45$ 
deg$^{-2}$ at a flux limit of $10^{-14}$ \cgs.  Within the reasonable 
hypothesis that the fraction of obscured QSOs among the sources with 
X/O$>$8 and log\,N$_H$(z=0)$>21$ remains constant down to fluxes of 
$10^{-15}$ \cgs, an obscured QSOs density of $\sim350$ deg$^{-2}$ is 
implied. 
A much more conservative estimate of the obscured QSO space density 
($\sim100$ deg$^{-2}$ at $10^{-15}$ \cgs), which should probably be 
considered as a lower limit, is obtained assuming that all sources 
with optical counterparts fainter than R=25.5 have X--ray luminosities 
below the highly obscured QSO threshold. 
 
The three black solid curves in figure \ref{xolx}b) superimposed to 
the number counts of high-luminosity, highly obscured objects from 
both HELLAS2XMM and CDFS+CDFN samples are the number counts of 
obscured QSO predicted by the Comastri et al. (1995, 2001) XRB 
synthesis models.  These models assume an evolution of the X--ray 
luminosity function parameterized by a pure luminosity evolution (PLE) 
law [$L(z) \propto L(z=0) \times (1+z)^{2.6}$] up to $z$ = 1.5 and 
constant up to a maximum redshift z$_{max}$ (z$_{max}$=4,5,7; see 
figure \ref{xolx}b).  These predictions (depending also on the adopted 
maximum redshift) lie in between the two number counts obtained 
assuming that 70\% of the high X/O sources are obscured QSOs or 
including in the obscured QSO sample sources with optical counterparts 
brighter than R=25.5 only, respectively.  The red curve in the figure 
\ref{xolx}b) represents the prediction obtained by the luminosity 
dependent density evolution (LDDE) model of the hard X--ray luminosity 
function described in La Franca et al. (2005).  Not surprisingly, the 
shape of the number counts relations obtained by integrating the PLE 
and the LDDE models are different.  The latter slightly overpredicts 
the highly obscured QSO number counts at fluxes $\gs 10^{-14}$ \cgs, 
while at lower fluxes agrees pretty well with the expectations based 
on the identification of obscured QSOs with 70\% of the high X/O 
sources (see Section 3.1). The same model predicts that the number of 
obscured QSO with logL(2-10keV)$>44$ is comparable to that of 
unobscured QSO. A similar conclusion was found by Perola et al. (2004) 
using a smaller source sample (the HELLAS2XMM 1dF sample).  However, 
most of these hard X--ray selected highly obscured QSOs have column 
densities in the range $N_H \sim 10^{22-23}$ cm$^{-2}$ with only a 
handful of the faintest sources which could be Compton thick (N$_H$ 
$>$~10$^{24}$ cm$^{-2}$).  According to the most recent version of the 
AGN synthesis models for the X--ray background (Gilli et al. 2006), 
the fraction of Compton-thick AGN in deep XMM--{\it Newton} and {\it 
Chandra} surveys is expected to be at most a few percent.  Indeed, 
Tozzi et al. (2006) find that only about 4 \% of the sources in the 
CDFS have a 0.5--8 keV spectrum consistent with Compton-thick 
absorption and/or pure reflection.  Therefore, we still may be viewing 
just the tip of the iceberg of highly obscured sources at 
high--redshift. 
 
An approach to find highly obscured, Compton-thick, QSOs is to select 
sources with QSO luminosities in the mid--infrared band and faint or 
extremely faint near--infrared and optical counterparts.  
Martinez-Sansigre et al. (2005) estimate that probably more than half 
of the high-luminosity QSOs are highly obscured, although with large 
uncertainties. Unfortunately, the X--ray properties of infrared 
selected sources are not known, and therefore it is difficult to 
understand how the mid--infrared selection compares with the X--ray 
one. In particular, it is not clear what is the fraction of the 
mid--infrared selected highly obscured QSOs which would have been 
selected by hard X--ray surveys, and whether the mid-infrared 
selection is truly more efficient in discovering Compton-thick 
sources.  To answer to these questions we can take advantage of 
the study of fields with both X--ray and mid--infrared coverage; for 
example, we have under analysis Spitzer observations of 12 HELLAS2XMM 
sources (Pozzi et al.  in preparation). Furthermore, the study of the 
ELAIS-S1 (Puccetti et al. 2006) and the COSMOS fields, and/or deep 
X--ray follow-up observations of the mid--infrared selected sources in 
the Spitzer First Look Survey (Martinez-Sansigre et al. 2005) will 
certainly bring new light on this topic. 

\subsection{XBONGs}  
At the other extreme of the X--ray-to-optical flux ratio distribution, 
we find the other class of ``elusive'' AGN (i.e., XBONGs), objects 
which have Seyfert--like X--ray luminosities but without any signature 
of nuclear activity in either optical imaging or spectroscopy 
(Comastri et al. 2002; Moran et al. 2002; Severgnini et al. 2003). In 
order to provide a quantitative definition of these objects, the flux 
and luminosity of the [OIII]$\lambda5007$ emission (or their upper 
limits) are measured for the 59 spectra in the full HELLAS2XMM sample 
that cover this transition.   The HELLAS2XMM AGNs with [OIII] 
detection have a $L_{2-10 keV}/L_{[OIII]}$ ratio between 3 and 1000 
with a logarithmic median and interquartile range of 2.14$\pm$0.38. 
This is systematically higher than optically selected sources. As an 
example, unobscured PG quasars at z$<$0.4 have $L_{2-10 
keV}/L_{[OIII]}$ in the range 6-500 (Laor et al. 1997), Compton thin, 
Seyfert 2 galaxies have $L_{2-10 keV}/L_{[OIII]}$ between 3 and 50 
while Compton-thick Seyfert 2 galaxies have $L_{2-10 keV}/L_{[OIII]}$ 
between 0.03 and 1 (Maiolino et al. 1998, Bassani et al. 1999). All 
the HELLAS2XMM AGNs have $L_{2-10 keV}/L_{[OIII]}>3$ and therefore, as 
discussed above, no Compton-thick object (log$N_H \gs 24$) is likely 
to be present in the HELLAS2XMM sample.  The discrepancy between the 
$L_{2-10 keV}/L_{[OIII]}$ ratio of the HELLAS2XMM AGN and optically 
selected AGNs could be due to the fact that the [OIII] luminosity of 
the HELLAS2XMM AGN is not corrected for reddening. This correction can 
be very large for highly obscured sources. Indeed, the typical 
correction in Bassani et al. (1999) is a factor between 2 and 10, with 
a few extreme objects with a correction approaching 50.  The 
logarithmic median $L_{2-10 keV}/L_{[OIII]}$ and its interquartile 
range of a sample of 24 Compton thin, narrow line AGN from the 
catalogs of Maiolino \& Rieke (1995) and Bassani et al. (1999), 
obtained without correcting for extinction, plus 17 PG quasar from the 
Laor et al (1997) sample (41 sources in total), is 1.69$\pm$0.30, 
lower than the median of the HELLAS2XMM sources.  The probability that 
the two log$L_{2-10 keV}/L_{[OIII]}$ distributions are drawn from the 
same parent population is 1.8\%, using the Kolmogorov-Smirnov test. 
The higher $L_{2-10keV}/L_{[OIII]}$ ratio of X-ray selected AGN with 
respect to optically selected AGN is intriguing, because it suggests 
that the latter samples are at least partly incomplete, and that 
[OIII] emission is not a perfect isotropic indicator of the nuclear 
power. However, this must be confirmed by using larger samples with 
both good X-ray and [OIII] determinations. 
 
Narrow-line AGNs tend to have higher column densities than broad-line 
AGNs, but we do not see any correlation between the column density and 
$L_{2-10 keV}/L_{[OIII]}$ (see figure \ref{oiii}b). On the other hand, 
XBONG candidates  have lower [OIII] fluxes compared with the other 
sources with similiar X--ray flux 
(see figure \ref{oiii}a) and have $L_{2-10 keV}/L_{[OIII]}\gs1000$ 
(see figure \ref{oiii}b). This suggests that the $L_{2-10 
keV}/L_{[OIII]}$ ratio is a robust index to define X-ray bright but 
optically normal galaxies (XBONGs), at least in the X-ray and [OIII] 
flux ranges covered by the present observations. Note that this 
classification criterion is nearly independent by the galaxy dilution 
of the nuclear spectrum, which has been suggested to be a key 
ingredient in the classification of XBONGs (see Severgnini et 
al. 2003, Moran et al.  2002, Georgantopoulos \& Georgakakis 2005), 
since it is related to a direct measurement of the line flux and not 
to its equivalent width.  Indeed, the main issue about the XBONGs 
nature is whether they are just typical AGNs (in terms of both X--ray 
and line emission luminosity) or, rather, they represent a truly 
distinct class. Our results strongly point towards the latter 
hypothesis. 
   
To further investigate this issue, we compared the 7 XBONG candidates 
to 7 narrow-line AGNs found in the same redshift interval 
(0.075--0.32).  XBONGs and narrow-line AGNs have very similar X--ray 
luminosity (median log\,L(2-10 keV)= 42.94 vs. 42.76), optical 
luminosity (median L(R)= 10.60 vs. 10.77 L$_{\odot}$), 
X--ray--to--optical flux ratio (median X/O= 0.35 vs. 0.26) and absorbing 
column densities (median log\,N$_H$ = 21.8 vs. 21.6).  On the other 
hand, while the narrow-line AGNs have median [OIII]$\lambda5007$ flux 
and luminosity of $9.2\times10^{-16}$ \cgs and $1.6\times10^{41}$ erg 
s$^{-1}$, XBONGs have 3$\sigma$ upper limits in the ranges 
$1.5-9.4\times10^{-17}$ \cgs and $2.5\times10^{38}-1.4\times10^{40}$ 
erg s$^{-1}$, i.e., 10--50 times lower than narrow-line AGNs.  The 
above findings suggest that while the central engine of narrow-line 
AGNs and XBONGs is likely to be the same, narrow emission lines in 
XBONGs are strongly inhibited or obscured. 
 
At least three possibilities are envisaged to explain the lack of 
optical line emission: 
\begin{itemize}  
 
\item 1) the physical size of the narrow emission-line region could be 
reduced with respect to normal AGNs, or even absent (see, e.g., 
Hawkins 2004) 
 
\item 2) the optical-UV nuclear continuum could be screened by 
circumnuclear absorbing gas and dust covering a large solid angle at 
the nuclear engine, so that the ionizing radiation does not (or only 
partially) reach the narrow emission-line clouds; 
 
\item 3) the emission lines could be quenched by obscuring dust 
spatially distributed on larger scales (i.e., kpc dust lanes) as 
observed in HST images of nearby Seyfert 2 galaxies (Malkan et 
al. 1998) 
 
\end{itemize}  
 
To disentangle between these hypotheses, we are pursuing an intensive 
multiwavelength observing strategy, including further high angular 
resolution X--ray imaging with {\it Chandra}, and spatially resolved 
optical spectroscopy with the VIMOS Integral Field Unit at VLT. 
 
\section{Conclusions} 
 
We have obtained optical photometry and spectroscopy for a sample of 
110 sources detected in the 2--10 keV band in five additional 
XMM--{\it Newton} fields of the  HELLAS2XMM serendipitous survey, 
covering additional 0.5 deg$^2$ of the sky  at the bright flux limit (see Table 2). 
  We report the 
spectroscopic identification of 59 new redshifts, bringing to 159 the 
total number of X--ray sources with redshift identifications in the 
full HELLAS2XMM sample (including the three near-infrared 
spectroscopic redshifts reported by Maiolino et al. 2006). 
 
Combining the redshift information of the HELLAS2XMM sample with that 
available in the CDFS and CDFN surveys, the fraction of highly obscured 
type 2 QSOs is estimated to be of the order of  $\sim$~13\% of the total, 
rather independent from the X--ray flux for F(2-10 keV)$>10^{-15}$\cgs.  
This result is in rough agreement with the Comastri et al. (2001) and La 
Franca et al. (2005) models and implies that about half of the luminous 
AGNs (logL(2-10keV)$>44$) are obscured. 
 
We find that the 7 XBONG candidates in the full HELLAS2XMM sample have 
$L_{2-10 keV}/L_{[OIII]}\gs1000$, while all but one of the other 
HELLAS2XMM AGNs have an X--ray-to-[OIII] ratio well below this 
value.\\  
 Although the sample is very small and, therefore, subject to large uncertainties,  
our results suggest that this ratio can be used to efficiently 
select  XBONG candidates. Multiwavelenght works with further surveys are needed to 
strengthen this result. 
 
\bigskip 
 
This research has been partially supported by ASI, INAF--PRIN 270/2003 
and MIUR Cofin-03-02-23 grants. We acknowledge an anonymous referee for comments that 
improved the presentation. We thanks L. Pentericci, P. Tozzi and P. Severgnini for useful 
discussions.

\newpage

\setcounter{table}{0}
\begin{landscape} 
\begin{table}[ht] 
\caption{\bf The HELLAS2XMM second source sample} 
\begin{tabular}{lcccccccccc} 
\hline 
\hline 
Observation Id & X-ray Position& Optical Position&  Offaxis             & $\Delta$ &Prob & F(2-10 keV) & R & Class. & z & log\,L$_{2-10 keV}$ \\  
         &  (J2000)          & (J2000)        &   (arcmin)                   & (arcsec) & &($10^{-14}$ cgs) & &     &    & (erg/s) \\         
\hline 
H2XMMJ140144.9+025332$^a$ & 14 01 44.9 +02 53 32 & 14 01 45.0 +02 53 34 &      11.0                 &  2.8 &0.9990& 5.13  &  18.00 & ETG & 0.2506 & 42.97 \\ 
                              &      & 14 01 44.9 +02 53 25 &                                  &  6.6 &0.9171& --  & 20.69 & -- & -- & -- \\ 
H2XMMJ140144.8+024845 & 14 01 44.8  +02 48  45 & 14 01 44.8  +02 48  40 &   11.8                   &  4.8 && 1.86  &  24.50 & -- & --  & -- \\ 
H2XMMJ140139.6+025722 & 14 01 39.6  +02 57  22 & 14 01 39.5  +02 57  21 &    10.6                  &  1.5 && 2.44  &  19.10 & AGN1 & 0.2490 &   42.64 \\ 
H2XMMJ140137.1+024604 & 14 01 37.1  +02 46  04 & 14 01 37.3  +02 46  06 &    11.5                  &  4.3 && 2.19  &  20.07 & --  & --  & -- \\ 
H2XMMJ140132.1+025222 & 14 01 32.1  +02 52  22 &  --      &                  7.9                  &   -- & &	 1.08  & $\gs$23.0  & --   &  --  & -- \\ 
H2XMMJ140130.8+024532$^a$ & 14 01 30.8 +02 45  32 & 14 01 30.6 +02 45  31 &    10.7                &  3.1 &0.9808& 4.08  &  20.93 & ELG &  0.7456 &   43.97 \\ 
                          &                       &  14 01 31.1 +02 45  31 &                   &  4.8 &0.7087& --  & 23.55  & -- &  -- &   -- \\ 
H2XMMJ140127.7+025607 & 14 01 27.7 +02 56 07 & 14 01 27.7 +02 56  06     &     7.4                &  0.7 && 57.0  &  19.70 & AGN1 &  0.2645 &   44.07 \\ 
H2XMMJ140125.3+024620$^a$ & 14 01 25.3 +02 46  20 & 14 01 25.2 +02 46  21 &    9.2                &  1.2 &0.9970& 1.29  &  20.99 & AGN2 &  0.4332 &   42.91 \\ 
                          &                        & 14 01 25.3  +02 46 19 &                   &  1.6 &0.9493& --  & 23.94  & -- &  -- &   -- \\ 
H2XMMJ140117.5+024349$^a$ & 14 01 17.5 +02 43  49 & 14 01 17.5 +02 43  51 &    10.2                &  2.1 &0.9979& 2.70  &  19.60 & AGN1 &  0.3630 &   43.06 \\ 
                            &                   &14 01 17.5 +02 43 48 &                        &  1.0 &0.9866& --  & 23  & -- &  -- &   -- \\ 
H2XMMJ140115.0+024821 & 14 01 15.0 +02 48  21 & 14 01 15.2 +02 48  19     &     6.0               &  3.4 && 1.66  &  19.75 &AGN1  &  1.5229 &   44.31 \\ 
H2XMMJ140109.9+024339$^a$ & 14 01  09.9 +02 43  39 & 14 01  09.9 +02 43  39 &   9.7               &  0.2 &1& 1.53  &  20.75 & AGN1 &  1.3550 & 44.15 \\ 
                                        &          &14 01 09.9 +02 43 37&                      &  2.4 &0.9664 & --  &   & ETG & 0.9356 &   -- \\ 
H2XMMJ140109.0+025651 & 14 01 09.0 +02 56  51 & 14 01 08.8 +02 56  50 &     4.3                    &  3.7 & &	 2.48  &  22.57 & AGN1 &  1.8330 & 44.67 \\ 
H2XMMJ140057.3+023942 & 14 00 57.3 +02 39  42 & 14 00 57.4 +02 39  43 &     13.5               &  1.6 & &	 3.64  &  19.50 & ETG &  0.2650 & 42.88 \\ 
H2XMMJ140053.1+024150 & 14 00 53.1 +02 41  50 & 14 00 53.0 +02 41  51 &        11.4                &  1.3 & &	 2.07  &  23.61 & --   & --  & -- \\ 
H2XMMJ140053.1+030104$^a$ & 14 00 53.1 +03 01   04 & 14 00 53.1 +03 01   06 &  8.2                &  1.3 &0.9985& 	 5.02  &  20.18 & AGN1 &  1.3045 & 44.63 \\ 
                             &                     &14 00 53.2 +03 01 10 &                     &  5.9 &0.6339& --  & 22.80  & -- &  -- &   -- \\ 
H2XMMJ140051.1+025906 & 14 00 51.1 +02 59  06 & 14 00 51.4  +02 59  05 &      6.4                 &  4.1 & &	 2.14  &  18.99 &  AGN2 &  0.2564 & 42.61 \\ 
H2XMMJ140049.1+025850 & 14 00 49.1 +02 58  50 & 14 00 49.0  +02 58  53 &      6.4                 &  3.5 & &	 1.41  &  21.23 & AGN1$^b$ &  1.8222 & 44.42 \\ 
H2XMMJ140040.9+025353 & 14 00 40.9 +02 53  53 & 14 00 40.8  +02 53  53 &      5.0                 &  1.4 & &	 1.75  &  24.19 & --   & --  & -- \\ 
H2XMMJ140038.7+024322 & 14 00 38.7 +02 43  22 & 14 00 38.6  +02 43  23 &       11.2                &  1.9 & &	 2.51  &  20.72 &AGN1  &  0.6634 &   43.63 \\ 
H2XMMJ140033.3+025810 & 14 00 33.3 +02 58  10 & 14 00 33.4  +02 58  13 &      8.5                 &  3.4 & &	 2.05  &  22.87 & AGN1 &  1.0260 &   44.00 \\ 
H2XMMJ140033.0+025740 & 14 00 33.0 +02 57  40 & 14 00 33.3  +02 57  42 &      8.3                &  5.0 & &	 2.17  &  23.75 & -- &  --    &   --      \\ 
H2XMMJ140019.3+025638 & 14 00 19.3 +02 56  38 & 14 00 19.1  +02 56  40 &      10.9                 &  4.0 & &	 2.33  &  22.52 & AGN1 &  1.1115 &   44.13 \\ 
H2XMMJ133807.5+242412 & 13 38 07.5 +24 24  12 & 13 38 07.5 +24 24  11 &        11.3                 &  0.7 &&  11.9 &  18.30 &AGN1  &  0.6336 &  44.27 \\ 
H2XMMJ133712.7+243251 & 13 37 12.7 +24 32  51 & 13 37 12.8 +24 32  54 &        9.9                   &  3.6 & & 3.89  &  19.60 & --   & --  & -- \\ 
H2XMMJ133643.1+242646 & 13 36 43.1 +24 26  46 &            --     &            8.8                   &  --  &&  3.46  & $\gs23.0$& --& --      &  --  \\ 
H2XMMJ133630.2+242625 & 13 36 30.2 +24 26  25 & 13 36 30.0 +24 26  22 &        11.4                  &  4.4 &&  2.97  &  18.60 & ELG &  0.2551 &   42.75\\ 
H2XMMJ133702.2+242434 & 13 37 02.2 +24 24  34 & 13 37 02.2 +24 24  36 &        3.9                   &  1.7 & & 1.56  &  21.15 & --    & --  & -- \\ 
H2XMMJ133730.6+242306 & 13 37 30.6 +24 23  06 & 13 37 30.8 +24 23  05 &        2.8                   &  3.5 & & 3.19  &  20.50 &  AGN1 &  1.2798 & 44.42\\ 
H2XMMJ133717.9+242148$^a$ & 13 37 17.9 +24 21  48 & 13 37 17.9 +24 21  49 &    1.5                   &  1.5 &0.9905& 	 1.72  &  22.07 & AGN2 &  0.3431 & 42.81\\ 
                                              &   & 13 37 18.0  +24 21 44 &                    &  4.5 &0.8867& --  & 22.58  & -- &  -- &   -- \\ 
H2XMMJ133649.4+242004 & 13 36 49.4  +24 20  04 & 13 36 49.3  +24 20  00 &      7.2                 &  3.3 & & 3.86  &  20.03 & --    & --  & -- \\ 
H2XMMJ133714.0+241960 & 13 37 14.0  +24 19  60 & 13 37 14.1  +24 20  01 &      3.2                 &  1.7 & & 1.61  &  21.16 & AGN1 &  1.5521 &  44.32\\ 
H2XMMJ133637.4+241935 & 13 36 37.4  +24 19  35 &                    --  &      9.9                 &  -- & &	 2.85  & $\gs$24.20 &  --  &  --  & -- \\ 
H2XMMJ133749.2+241942 & 13 37 49.2  +24 19  42 & 13 37 49.5 +24 19  42 &       7.8                 &  4.1 &&  1.68  &  23.05 & --    & --  & -- \\ 
H2XMMJ133724.3+241922 & 13 37 24.3  +24 19  22 & 13 37 24.5 +24 19  23 &       3.9                 &  2.8 &&  1.48  &  23.29 & --    & --  & -- \\ 
H2XMMJ133659.3+241916$^a$ & 13 36 59.3 +24 19  16 & 13 36 59.4 +24 19  15 &    5.7                 &   1.3 &0.9953&  2.30  &  21.75 & ELG  &  0.5008 & 43.31 \\ 
                         &                       & 13 36 59.0  +24 19 11 &                     &   5.6 &0.9572& --  & 20.7  & -- &  -- &   -- \\ 
\hline  
\end{tabular}  
\end{table}  
\end{landscape}  
  
\setcounter{table}{0} 
\begin{landscape}     
\begin{table}[ht]  
\caption{\bf The HELLAS2XMM second source sample, continue}  
\begin{tabular}{lccccccccccc}  
\hline  
\hline  
Id & X-ray Position& Optical Position&     Offaxis              &      $\Delta$& Prob & F(2-10 keV) & R & Class. & z & log\,L$_{2-10 keV}$ \\  
   &  (J2000)         & (J2000)        &    (arcmin)                    &(arcsec) && ($10^{-14}$ cgs) & &     &    & (erg/s) \\         
\hline  
H2XMMJ125605.6+220719  & 12 56 05.6  +22 07  19 & 12 56 05.7  +22 07  19 &         13.1    &               1.8 & & 4.74  &  20.07 & AGN1 &  1.1450 & 44.47 \\ 
H2XMMJ125625.6+220717  & 12 56 25.6  +22 07  17 & 12 56 25.7  +22 07  20 &         9.1     &               3.3 & & 2.17  &  20.55 &AGN1  &  1.7310 & 44.56 \\ 
H2XMMJ125719.2+220030  & 12 57 19.2  +22 00  30 & 12 57 19.4  +22 00  32 &         5.2     &               3.4 & & 1.19  &  21.07 & AGN1 &  1.5060 &  44.15\\ 
H2XMMJ125704.3+220037  & 12 57 04.3  +22 00  37 & 12 57 04.1  +22 00  40 &         2.1     &               4.3 & & 0.83 &  23.44 & --    & --  & -- \\ 
H2XMMJ125647.4+215946 & 12 56 47.4  +21 59  46 & 12 56 47.2  +21 59  47 &          3.2     &               2.6 && 1.67  &  22.18 & --    & --  & -- \\ 
H2XMMJ125602.8+215952 & 12 56 02.8  +21 59  52 & 12 56 02.7  +21 59  52 &          12.8    &                1.4 && 11.1 &  19.10 & --    & --  & -- \\ 
H2XMMJ125750.6+215936 & 12 57 50.6  +21 59  36 & 12 57 50.5  +21 59  34 &          12.5    &               2.2 && 2.55  &  18.90 & AGN1 &  0.6150 & 43.56\\ 
H2XMMJ125632.8+215936$^a$ & 12 56 32.8 +21 59  36 & 12 56 32.8 +21 59  37 &        6.2     &               1.1 &0.9989& 3.07  &  20.10 &AGN1$^c$ & 0.5160 & 43.47\\ 
                       &                       & 12 56 33.2  +21 59 37 &                   &         5.8 &0.9341& --  & 21.26  & ELG & 0.2790 &   -- \\ 
H2XMMJ125709.0+215800 & 12 57  09.0 +21 58  00 & 12 57  08.9  +21 57  57 &        4.8      &                3.1 && 0.67 &  20.79 & AGN1 & 0.9112 & 43.39\\ 
H2XMMJ125629.4+215704 & 12 56 29.4 +21 57  04 & 12 56 29.3  +21 57   06 &         8.1      &                 2.3 & &1.27  &  22.41 & AGN1$^d$ & 0.6346 &43.30\\
H2XMMJ125732.6+215708 & 12 57 32.6 +21 57  08 & 12 57 32.7  +21 57   08 &         9.5      &                0.9 & &1.39  &  19.40 & --   & --  & -- \\ 
H2XMMJ125638.9+215625 & 12 56 38.9 +21 56  25 & 12 56 38.9  +21 56  26 &          7.0      &                1.0 & &1.01  &  22.64 & ELG  &  0.4703 & 42.89 \\ 
H2XMMJ125712.2+215523 & 12 57 12.2 +21 55 23 & 12 57 12.1  +21 55 22 &            7.4      &               1.6& &1.14  &  20.78 & AGN1 &  0.8000 &   43.49 \\ 
H2XMMJ125637.3+215439 & 12 56 37.3 +21 54 39 & 12 56 37.0  +21 54 36 &            8.7      &                5.2& &1.08  &  23.71 & --   & --  & -- \\ 
H2XMMJ125650.6+215458$^a$ & 12 56 50.6 +21 54  58 & 12 56 50.6 +21 54  59 &       7.2      &                0.7 & 0.9996&	 2.65  &  20.00 & AGN1 &  0.6450 &   43.63 \\ 
                                  &               &12 56 50.9  +21 54 57 &                 &         4.7 &0.7478 &--  & 22.81  & -- & -- &   -- \\ 
H2XMMJ125706.3+215506$^a$ & 12 57 06.3 +21 55 06 & 12 57 06.2 +21 55  06 &      7.2        &            1.4 &0.9737 &	 0.99 &  22.85 & AGN1 &  1.8750 &   44.30 \\ 
                                        &          &12 57 06.1  +21 55 04 &                &         3.5 &0.9292& --  & 22.77  & -- & -- &   -- \\ 
H2XMMJ125715.9+215432  & 12 57 15.9  +21 54  32 & 12 57 15.7 +21 54  31 &       8.5        &               2.6 & &1.82  &  19.32 & AGN1 &  0.4011 &   42.99 \\ 
H2XMMJ125712.2+215358  & 12 57 12.2  +21 53  58 & 12 57 12.3 +21 53  59 &       8.7        &                   1.8 & & 1.66  &  20.71 & ELG  &  0.6854 &   43.49 \\ 
H2XMMJ125640.3+215351  & 12 56 40.3  +21 53  51 & 12 56 40.4 +21 53  51 &       9.0        &               1.5 & & 1.17  &  22.53 & --   & --  & -- \\ 
H2XMMJ125628.1+215405  & 12 56 28.1  +21 54  05 & 12 56 27.9 +21 54  06 &       10.4       &                 2.6 & & 5.58  &  19.69 & AGN1 &   1.8670 &   45.04 \\ 
H2XMMJ125710.6+215353  & 12 57 10.6  +21 53  53 & 12 57 10.2 +21 53  53 &       8.6        &                5.7 & & 2.06  &  19.92 & AGN1 &  0.7977 &   43.74 \\ 
H2XMMJ125650.3+215329  & 12 56 50.3  +21 53  29 & 12 56 50.5 +21 53  30 &       8.6        &               3.3 & & 1.25  &  18.69 & ETG$^e$ &  0.3991 &   42.82 \\ 
H2XMMJ125659.0+215347$^b$ & 12 56 59.0 +21 53  47 & 12 56 58.8 +21 53  46 &     8.2        &                    3.5& & 	 6.06  &  17.42 & AGN2 &  0.1871 &   42.77 \\ 
H2XMMJ125654.1+215318  & 12 56 54.1 +21 53  18 & 12 56 54.1 +21 53  17 &        8.7        &                   1.0 & & 2.44  &  20.76 & AGN2 &  0.9093 &   43.95 \\ 
H2XMMJ125653.7+215125$^a$ & 12 56 53.7 +21 51  25 & 12 56 53.6 +21 51  26 &     10.6       &                    2.1 &0.9936 &	 5.17  &  20.47 & AGN1$^f$ &  0.8940 &   44.26 \\ 
                              &                      &12 56 53.4 +21 51 24 &               &              4.7 &0.8972& --  & 22.34  & AGN2 & 0.9127 &   -- \\ 
H2XMMJ125633.1+215147$^a$ & 12 56 33.1 +21 51  47 & 12 56 32.9 +21 51  52 &     11.6       &                   4.9 &0.9882&  2.95  &  19.60 & AGN1$^c$ &   0.8950 &   44.01 \\ 
                                         &        &12 56 33.0 +21 51 49 &                  &      2.5 &0.9856& --  & 21.36  & ETG & 0.7533 &   -- \\ 
H2XMMJ110522.5+381103 & 11 05 22.5  +38 11  03 & 11 05 22.6  +38 11  05 &     11.8         &            2.1 &&  2.19  &  19.96 & AGN1 &  1.2585 &   44.23 \\ 
H2XMMJ110522.0+381401 & 11 05 22.0  +38 14  01 & 11 05 22.0  +38 14  02 &     12.1         &            1.1 &&  5.86  &  11.30 & AGN2 &  0.0276 &   41.05 \\ 
H2XMMJ110517.9+381051 & 11 05 17.9  +38 10  51 & 11 05 17.8  +38 10  51 &     10.9         &            1.7 &&  2.24  &  22.17 & --    & --  & -- \\ 
H2XMMJ110512.3+382129 & 11 05 12.3  +38 21  29 & 11 05 12.8  +38 21  29 &     14.4         &            5.8 &&  4.28  &  19.70 & --    & --  & -- \\ 
H2XMMJ110509.7+381253 & 11 05 09.7  +38 12  53 & 11 05 09.6  +38 12  55 &     9.5          &            2.7 &&  1.39  &  23.41 & --    & --  & -- \\ 
H2XMMJ110449.2+381810 & 11 04 49.2  +38 18  10 & 11 04 49.2  +38 18  11 &     9.1          &            1.7 &&  2.06  &  18.30 & AGN1  & 1.9451 &   44.65 \\ 
H2XMMJ110447.6+380407 & 11 04 47.6  +38 04  07 & 11 04 47.6  +38 04  07 &     8.3          &            0.6 & 	& 1.59  &  21.30 & AGN1  &  2.2670 &   44.69 \\ 
H2XMMJ110444.2+381449 & 11 04 44.2  +38 14  49 & 11 04 43.9  +38 14  48 &     5.9          &            4.3 &&  3.29  &   7.60 &  * &  0.0000 &   40.00 \\ 
H2XMMJ110438.2+382500 & 11 04 38.2  +38 25  00 & 11 04 38.1  +38 25  01 &     14.6         &            2.0 &&  1.63  &  22.45 & --    & --  & -- \\ 
H2XMMJ110437.8+382304 & 11 04 37.8  +38 23  04 & 11 04 37.6  +38 23  02 &     12.6         &            2.1 &&  1.57  &  22.06 & --    & --  & -- \\ 
H2XMMJ110435.2+382139 & 11 04 35.2  +38 21  39 &  --       &                  11.2         &            -- & &	 1.26  & $\gs$23.30 & --   &  --  & -- \\ 
H2XMMJ110431.5+380305 & 11 04 31.5  +38 03  05 &  --       &                  7.9          &                -- & &	 1.26  & $\gs$23.50 & --   &  --  & -- \\ 
H2XMMJ110424.9+380024 & 11 04 24.9  +38 00  24 &  --       &                  10.4         &                -- & &	 1.54  & $\gs$23.50 & --   &  --  & -- \\ 
\hline 
\end{tabular} 
 \end{table}  
\end{landscape}

\setcounter{table}{0}      
\begin{landscape}
\begin{table}[ht]  
\caption{\bf The HELLAS2XMM second source sample, continue}  
\begin{tabular}{lcccccccccccc}  
\hline  
\hline  
Id & X-ray Position & Optical Position&   Offaxis &            $\Delta$ &Prob & F(2-10 keV) & R & Class. & z & log\,L$_{2-10 keV}$ \\  
   &  (J2000)         & (J2000)              &    (arcmin)&              (arcsec)& & ($10^{-14}$ cgs) & &     &    & (erg/s) \\         
\hline  
H2XMMJ110420.8+380443 & 11 04 20.8  +38 04  43 & 11 04 20.9  +38 04  37 &     6.0                          &      6.1 & &	 1.28  &  23.34 & --   &  --  & -- \\ 
H2XMMJ110418.2+382047 & 11 04 18.2  +38 20  47 & 11 04 18.3  +38 20  47 &     10.1                         &      0.6 &&  2.66  &  22.10 & AGN1 &  2.0487 &   44.82 \\ 
H2XMMJ110416.2+380241 & 11 04 16.2  +38 02  41 & 11 04 16.0  +38 02  36 &     8.2                          &      5.6 & &	 2.08  &  20.52 &  --  &  --  & -- \\ 
H2XMMJ110414.7+380714 & 11 04 14.7  +38 07  14 & 11 04 14.8  +38 07  15 &     3.8                          &      0.9 && 	 5.70  &  22.75 & --   &  --  & -- \\ 
H2XMMJ110413.2+382203 & 11 04 13.2  +38 22  03 & 11 04 13.3  +38 22  07 &     11.4                         &      4.6 &&  1.15  &  21.00 & --   &  --  & -- \\ 
H2XMMJ110402.8+375950 & 11 04 02.8  +37 59  50 & 11 04 03.0  +37 59  50 &     11.6                         &      2.2 &&  3.75  &  20.52 &AGN1  &  0.2864 &   42.97 \\ 
H2XMMJ110345.8+380546 & 11 03 45.8  +38 05  46 & 11 03 46.3  +38 05  49 &     8.8                          &      6.0 & &	 2.46  &  22.20 & --   & --  & -- \\ 
H2XMMJ110343.6+381349 & 11 03 43.6  +38 13  49 & 11 03 43.6  +38 13  49 &     8.2                          &      0.5 & & 2.80  &  22.85 & --  &  --  & -- \\ 
H2XMMJ110339.9+380014 & 11 03 39.9  +38 00  14 & 11 03 40.0  +38 00  11 &     13.5                         &      3.5 & 	& 3.42  &  19.50 & AGN2  &  0.3078 &   43.00 \\ 
H2XMMJ110330.2+381607 & 11 03 30.2  +38 16  07 & 11 03 30.1  +38 16  06 &     11.6                         &      1.7 &&  1.35  &  22.75 & --   &  --  & -- \\ 
H2XMMJ110325.8+381212 & 11 03 25.8  +38 12  12 & 11 03 25.9  +38 12  11 &     11.3                         &      1.4 &&  5.68  &  14.40 & AGN2 &  0.0696 &   41.82 \\ 
H2XMMJ110322.6+380945 & 11 03 22.6  +38 09  45 & 11 03 22.8  +38 09  45 &     11.8                         &      1.6 & &	 2.46  &  21.85 & --   & --  & -- \\ 
H2XMMJ110318.6+381545 & 11 03 18.6  +38 15  45 & 11 03 18.4  +38 15  44 &     13.5                         &      2.6 & & 6.87  &  18.98 &AGN1   &  0.3140 &   43.32 \\ 
H2XMMJ110317.1+381336 & 11 03 17.1  +38 13  36 & 11 03 17.4  +38 13  38 &     13.2                         &      4.0 & & 3.66  &  19.51 & AGN1  &  1.2506 &   44.45 \\ 
H2XMMJ005019.4$-$515532 & 00 50 19.4 $-$51 55  32 & 00 50 19.5  $-$51 55  30 &            14.5             &      1.6 &&  2.50  &  21.60 & --    & --  & -- \\ 
H2XMMJ005026.3$-$515929 & 00 50 26.3 $-$51 59  29 & 00 50 26.3  $-$51 59  30 &           11.3              &        1.3 &&  1.89  &  22.30 & --    & --  & -- \\ 
H2XMMJ005009.4$-$515933 & 00 50 09.4 $-$51 59  33 & 00 50 09.7  $-$51 59  32 &            10.2             &         2.8 &&  4.38  &  20.05 & --    & --  & -- \\ 
H2XMMJ005030.6$-$520011 & 00 50 30.6 $-$52 00  11 & 00 50 30.8  $-$52 00  10 &           11.0              &         1.9 & & 13.5 &  18.78 & AGN1 &  0.4630 &   44.00 \\ 
H2XMMJ005030.7$-$520046$^g$ & 00 50 30.7 $-$52 00  46 & 00 50 30.9  $-$52 00  48 &       10.5              &        2.4& &  2.62  &  24.08 & AGN2 &  1.3553$^h$ &   44.39 \\ 
H2XMMJ005043.4$-$520116 & 00 50 43.4  $-$52 01  16 & 00 50 43.7  $-$52 01  17 &          11.4              &            3.1 & & 2.02  &  21.18 & --    & --  & -- \\ 
H2XMMJ005126.5$-$520220 & 00 51 26.5  $-$52 02  20 & 00 51 26.2  $-$52 02  21 &          16.3              &        3.0 &&  4.02  &  20.20 & --    & --  & -- \\ 
H2XMMJ005008.4$-$520350 & 00 50 08.4  $-$52 03  50 &  --     &                           6.1               &           -- & 	& 0.96 & $\gs$22.50 &  --   & --  & -- \\ 
H2XMMJ004953.1$-$520525 & 00 49 53.1  $-$52 05  25 & 00 49 52.9  $-$52 05  24 &          3.9               &         1.9 &&  0.95 &  21.65 & --    & --  & -- \\ 
H2XMMJ005031.6$-$520630$^g$ &00 50 31.6 $-$52 06  30 & 00 50 31.5 $-$52 06  34 &         6.9               &         4.1 &&  1.63  &  24.04 & --   & $>1.30^g$ & -- \\ 
H2XMMJ005017.1$-$520715 & 00 50 17.1  $-$52 07  15 & 00 50 17.4  $-$52 07  18 &          4.6               &          4.5 & & 0.71 &  23.40 & --   &  --  & -- \\ 
H2XMMJ005044.7$-$520735 & 00 50 44.7  $-$52 07  35 & 00 50 44.7  $-$52 07  36 &          8.5               &             0.9 & & 1.02  &  19.84 & --   &  --  & -- \\ 
H2XMMJ005012.3$-$520834 & 00 50 12.3  $-$52 08  34 & 00 50 12.3  $-$52 08  36 &          3.4               &             2.2 & & 0.61 &  22.28 & --   &  --  & -- \\ 
H2XMMJ004959.0$-$521112 & 00 49 59.0  $-$52 11  12 & 00 49 59.0  $-$52 11  10 &          2.3               &                1.4 & & 1.05  &  21.26 & --   &  --  & -- \\ 
H2XMMJ004955.7$-$521231 & 00 49 55.7  $-$52 12  31 &  --     &                           3.3               &              -- &&  0.76 & $\gs$22.50 & --  &  -- & -- \\ 
H2XMMJ004947.6$-$521249 & 00 49 47.6  $-$52 12  49 & 00 49 47.8  $-$52 12  48 &          3.5               &             1.9 & & 2.66  &  21.04 & AGN1  & 1.1046 &   44.19 \\ 
H2XMMJ004936.7$-$521306$^a$ & 00 49 36.7  $-$52 13 06 & 00 49 36.7 $-$52 13 07 &         4.3               &            0.3 & 0.9999&	 0.96 &  20.31 & AGN1  & 0.9290 &   43.57 \\ 
                              &                      &00 49 37.2 $-$52 13 07 &                             &       4.6 &0.9421& --  & 21.56  & -- & -- &   -- \\ 
H2XMMJ004950.9$-$521410  &00 49 50.9  $-$52 14  10 & 00 49 51.1  $-$52 14  13 &           4.8              &                2.9 & & 2.22  &  22.30 & --    & --  & -- \\ 
H2XMMJ004959.4$-$521412  &00 49 59.4  $-$52 14  12 & 00 49 59.2  $-$52 14  11 &           5.1              &             1.9 &&  1.40  &  21.70 & --    & --  & -- \\ 
H2XMMJ004935.2$-$521458  &00 49 35.2  $-$52 14  58 & 00 49 35.9  $-$52 14  56 &           6.1              &             6.7 &&  1.11  &  20.42 & --    & --  & -- \\ 
H2XMMJ005007.4$-$521508$^a$ & 00 50  07.4 $-$52 15 08 & 00 50 07.4  $-$52 15  08 &        6.3              &            0.6 &0.9998& 	 1.41  &  19.43 & AGN1$^b$  &  2.4220 &   44.71 \\ 
                                     &               &00 50 07.7 $-$52 15 06 &                             &       3.3 &0.9563& --  & 22.68  & -- & -- &   -- \\ 
H2XMMJ005032.1$-$521543  &00 50 32.1 $-$52 15  43 & 00 50 32.2  $-$52 15  43 &        9.0                  &                0.5 &&  6.41  &  19.77 & AGN1  &  1.2224 &   44.67 \\ 
\hline 
\end{tabular} 
 
Classification: AGN1; AGN2; ELG=Emission-Line Galaxy; ETG=Early-Type Galaxy; *=Star; 
$^a$ Two possible counterparts within $6''$; $^b$ Broad absorption line  QSO; 
$^c$ Two nearby objects, a type 1 AGN and an emission-line galaxy; 
$^d$ Tentative classification: low $S/N$ spectrum; 
$^e$ Extended, emission from a group  or  cluster of galaxies;  
$^f$ Two nearby objects, a type 1 AGN and a type 2 AGN; 
$^g$ Aperture photometry at the  position of a bright K source and photometric z  (Mignoli et al. 2004); 
$^h$ Near-infrared spectroscopy  (Maiolino et al. 2006).

\end{table} 
\end{landscape}

\setcounter{table}{2}

\begin{table*}[h!]   
\caption{\bf The HELLAS2XMM [OIII] sample}   
\begin{tabular}{lccccc}   
\hline   
\hline   
Observation Id	     &   z   &  Class. &logL$_{2-10 keV}$& LogF([OIII])& LogL([OIII]) \\   
             &       &         &     (erg/s)       &    (cgs) & (erg/s)       \\  
\hline   
H2XMMJ054034.2$-$283108 &     0.3794 &       AGN2  &  43.44   &       -15.44$^a$  &     41.12$^a$ \\   
H2XMMJ054024.7$-$284616 &     0.4844 &       AGN2  &  43.02   &       -16.35  &     40.43	\\   
H2XMMJ053959.0$-$283753 &      0.8420 &       AGN1  &  43.93   &       -14.96  &     42.31	\\   
H2XMMJ053929.5$-$284860 &     0.3171 &       AGN1  &  43.55   &       -15.32  &     41.08	\\   
H2XMMJ053910.7$-$283528 &       0.7700 &       AGN1  &   43.95  &       -14.64  &     42.55	\\   
H2XMMJ053850.9$-$283757 &      0.7630 &       AGN1  & 44.2    &       -14.82  &     42.36	\\   
H2XMMJ031416.4$-$764536 &     0.2456 &       AGN1  & 42.74    &        -15.3  &     40.87	\\   
H2XMMJ031200.4$-$770026 &      0.8410 &       AGN2  & 43.79    &       -16.59  &     40.68	\\   
H2XMMJ031136.0$-$765556 &      0.7090 &       ELG   & 43.49    &       -16.46  &     40.66	\\   
H2XMMJ031112.8$-$764706 &      0.7530 &       AGN1  &  43.72   &       -16.27  &      40.9	\\   
H2XMMJ031050.0$-$763904 &     0.3812 &       AGN1  &  44.12   &       -15.46$^a$  &  41.1$^a$\\   
H2XMMJ031037.4$-$764713 &      0.6410 &       ELG   &   43.43  &       -15.57  &     41.46	\\   
H2XMMJ030951.2$-$765825 &      0.2060 &       ELG   & 42.66    &       -15.87  &     40.14	\\   
H2XMMJ030912.1$-$765826 &     0.2651 &       AGN2  & 43.3    &       -15.03  &      41.2	\\   
H2XMMJ235933.4$-$250758 &      0.7380 &       AGN1  &  43.88   &       -15.63  &     41.52	\\   
H2XMMJ000102.4$-$245847 &     0.4331 &       AGN1  & 43.85    &       -16.22  &     40.46	\\   
H2XMMJ000100.2$-$250459 &     0.8504 &       AGN1  & 44.66    &       -15.04  &     42.24	\\   
H2XMMJ000036.6$-$250105 &     0.5917 &       AGN2  &43.64     &       -15.85  &     41.11	\\   
H2XMMJ000031.7$-$245459 &     0.2839 &       AGN1  & 43.35    &       -14.83  &     41.47	\\   
H2XMMJ000027.7$-$250441 &     0.3362 &       AGN1  & 43.43    &       -14.78  &     41.67	\\   
H2XMMJ000026.0$-$250648 &     0.4331 &       AGN1  & 43.03    &        -15.8  &     40.88	\\   
H2XMMJ003418.5$-$120809 &     0.2327 &       AGN2  & 43.04    &       -15.24  &     40.88	\\   
H2XMMJ204446.4$-$103839 &     0.6941 &       AGN2  & 43.84    &       -14.77  &     42.33	\\   
H2XMMJ204349.2$-$103746 &     0.5556 &       AGN1  & 43.82    &       -14.88  &     42.02	\\   
H2XMMJ140139.6+025722   &      0.2490 &       AGN1  &  42.64   &       -15.11  &     41.07	\\   
H2XMMJ140130.8+024532   &     0.7456 &       ELG   &  43.97   &       -15.74  &     41.42	\\   
H2XMMJ140125.3+024620   &     0.4332 &       AGN2  &  42.91   &        -14.36  &     42.32	\\   
H2XMMJ140117.5+024349   &      0.3630 &       AGN1  &  43.06   &       -15.2$^a$  & 41.32$^a$	\\   
H2XMMJ140051.1+025906   &     0.2564 &       AGN2  &  42.61   &       -14.53  &     41.67	\\   
H2XMMJ140038.7+024322   &     0.6634 &       AGN1  &  43.63   &       -15.78  &     41.28	\\   
H2XMMJ133630.2+242625   &     0.2551 &       ELG   &  42.75   &       -15.15  &     41.05	\\   
H2XMMJ133717.9+242148   &     0.3431 &       AGN2  &  42.81   &       -16.12  &     40.34	\\   
H2XMMJ133659.3+241916   &     0.5008 &       ELG   &  43.31   &       -15.89  &     40.92	\\   
H2XMMJ125632.8+215936   &      0.5160 &       AGN1  &  43.47   &       -15.78$^b$  & 41.06$^b$	\\   
H2XMMJ125709.0+215800   &     0.9112 &       AGN1  &  43.39   &       -16.04  &      41.3	\\   
H2XMMJ125629.4+215704   &     0.6346 &       AGN1  &  43.30   &       -16.51  &     40.51	\\   
H2XMMJ125638.9+215625 &     0.4703 &       ELG   &  42.89   &       -16.68  &     40.07	\\   
H2XMMJ125712.2+215523 &     0.8000 &       AGN1  &  43.49   &       -16.39  &     40.84	\\   
H2XMMJ125715.9+215432 &     0.4011 &       AGN1  &  42.99   &       -15.64  &     40.97	\\   
H2XMMJ125712.2+215358 &     0.6854 &       ELG   &  43.49   &       -16.59  &      40.5	\\   
H2XMMJ125710.6+215353 &     0.7977 &       AGN1  &  43.74   &       -15.88  &     41.34	\\   
H2XMMJ125659.0+215347 &     0.1871 &       AGN2  &   42.77  &       -14.56  &     41.37	\\   
H2XMMJ125654.1+215318 &     0.9093 &       AGN2  &  43.95   &       -15.59  &     41.75	\\   
H2XMMJ110522.0+381401 &    0.0276 &       ELG   &  41.05   &       -14.36  &     39.92	\\   
H2XMMJ110402.8+375950 &     0.2864 &       AGN1  &  42.97   &       -15.21  &     41.09	\\   
H2XMMJ110339.9+380014 &     0.3078 &       AGN2  & 43.00    &       -15.02  &     41.35	\\   
H2XMMJ110325.8+381212 &    0.0696 &       AGN2  &  41.82   &       -14.68  &     40.36	\\   
H2XMMJ110318.6+381545 &      0.3140 &       AGN1  &  43.32   &       -14.81  &     41.58	\\   
H2XMMJ005030.6$-$520011 &      0.4630 &       AGN1  &  44.00  &       -14.71  &     42.02	\\   
H2XMMJ053925.8$-$284456 &      0.0750 &       ETG    & 41.82   & $<-16.70$     &   $<38.41$ \\   
H2XMMJ031239.3$-$765133 &      0.1590 &       ETG    & 42.24   & $<-16.80$     &   $<38.98$ \\   
H2XMMJ031231.2$-$764324 &    0.0520 &       ETG$^c$& 41.08   & $<-16.26$     &   $<38.54$ \\   
H2XMMJ031124.8$-$770139 &     0.3199 &       ETG    & 42.93   & $<-16.42$     &   $<39.98$ \\   
H2XMMJ030952.2$-$764927 &     0.2049 &       ETG    & 42.22   & $<-16.74$     &   $<39.26$ \\   
H2XMMJ000030.1$-$251214 &     0.1543 &       ETG$^c$& 42.07   & $<-16.00$     &   $<39.75$ \\   
H2XMMJ204420.5$-$104904 &      0.3240 &       ETG    & 43.16   & $<-16.67$     &   $<39.75$ \\   
H2XMMJ140144.9+025332 &     0.2506 &       ETG    & 42.97   & $<-16.03$     &   $<40.16$ \\   
H2XMMJ140057.3+023942 &      0.2650 &       ETG    & 43.02   & $<-16.31$     &   $<39.92$ \\   
H2XMMJ125650.3+215329 &     0.3991 &       ETG$^c$& 42.82   & $<-17.18$     &   $<39.43$ \\   
\hline   
\end{tabular}   
 
$^a$ [OIII] lines within strong telluric absorption feature at $\sim$6900\AA\ ; 
$^b$ [OIII] lines within strong \newline telluric absorption feature at $\sim$7600\AA\ ; 
$^c$ Extended, emission from a group or cluster of galaxies;\newline  
$<$ indicates  $3\sigma$ upper limits.

\end{table*}   
  
\newpage 
 
\setcounter{table}{3}  
\begin{table*}[h!]   
\caption{\bf IAU names for the HELLAS2XMM source sample}   
\begin{tabular}{lcccccc}   
\hline   
\hline   
~~~~~~~~~~~~~~IAU name &   Id       &~~~~~~~~~IAU name&    Id       &~~~~~~~~~IAU name&  Id \\           
\hline  
\hline  
  
H2XMMJ053925.8$-$284456 &   05370024$^a$&  H2XMMJ133714.0$+$241960 &  13390131 & H2XMMJ005030.6$-$520011 &  16274066\\ 
H2XMMJ054034.2$-$283108 &   05370008$^a$&  H2XMMJ133637.4$+$241935 &  13390133 & H2XMMJ005030.7$-$520046 &  16274069\\ 
H2XMMJ054024.7$-$284616 &   05370135$^a$&  H2XMMJ133749.2$+$241942 &  13390134 & H2XMMJ005043.4$-$520116 &  16274078\\ 
H2XMMJ053959.0$-$283753 &   05370007$^a$&  H2XMMJ133724.3$+$241922 &  13390137 & H2XMMJ005126.5$-$520220 &  16274097\\ 
H2XMMJ053929.5$-$284860 &   05370003$^a$&  H2XMMJ133659.3$+$241916 &  13390140 & H2XMMJ005008.4$-$520350 &  16274117\\ 
H2XMMJ053910.7$-$283528 &   05370009$^a$&  H2XMMJ125605.6$+$220719 &  15300065 & H2XMMJ004953.1$-$520525 &  16274158\\ 
H2XMMJ053850.9$-$283757 &   05370020$^a$&  H2XMMJ125625.6$+$220717 &  15300067 & H2XMMJ005031.6$-$520630 &  16274181\\ 
H2XMMJ031239.3$-$765133 &   03120018$^a$&  H2XMMJ125719.2$+$220030 &  15300142 & H2XMMJ005017.1$-$520715 &  16274197\\ 
H2XMMJ031231.2$-$764324 &   03120008$^a$&  H2XMMJ125704.3$+$220037 &  15300143 & H2XMMJ005044.7$-$520735 &  16274212\\ 
H2XMMJ031124.8$-$770139 &   03120017$^a$&  H2XMMJ125647.4$+$215946 &  15300148 & H2XMMJ005012.3$-$520834 &  16274235\\ 
H2XMMJ030952.2$-$764927 &   03120501$^a$&  H2XMMJ125602.8$+$215952 &  15300151 & H2XMMJ004959.0$-$521112 &  16274292\\ 
H2XMMJ031416.4$-$764536 &   03120010$^a$&  H2XMMJ125750.6$+$215936 &  15300152 & H2XMMJ004955.7$-$521231 &  16274307\\ 
H2XMMJ031200.4$-$770026 &   03120016$^a$&  H2XMMJ125632.8$+$215936 &  15300153 & H2XMMJ004947.6$-$521249 &  16274314\\ 
H2XMMJ031136.0$-$765556 &   03120181$^a$&  H2XMMJ125709.0$+$215800 &  15300176 & H2XMMJ004936.7$-$521306 &  16274320\\ 
H2XMMJ031112.8$-$764706 &   03120011$^a$&  H2XMMJ125629.4$+$215704 &  15300186 & H2XMMJ004950.9$-$521410 &  16274339\\ 
H2XMMJ031050.0$-$763904 &   03120007$^a$&  H2XMMJ125732.6$+$215708 &  15300189 & H2XMMJ004959.4$-$521412 &  16274340\\ 
H2XMMJ031037.4$-$764713 &   03120028$^a$&  H2XMMJ125638.9$+$215625 &  15300193 & H2XMMJ004935.2$-$521458 &  16274349\\ 
H2XMMJ030951.2$-$765825 &   03120014$^a$&  H2XMMJ125712.2$+$215523 &  15300203 & H2XMMJ005007.4$-$521508 &  16274351\\ 
H2XMMJ030912.1$-$765826 &   03120034$^a$&  H2XMMJ125637.3$+$215439 &  15300207 & H2XMMJ005032.1$-$521543 &  16274363\\ 
H2XMMJ000030.1$-$251214 &   26900013$^a$&  H2XMMJ125650.6$+$215458 &  15300208 &&\\  
H2XMMJ235933.4$-$250758 &   26900028$^a$&  H2XMMJ125706.3$+$215506 &  15300209 &&\\ 
H2XMMJ000102.4$-$245847 &   26900003$^a$&  H2XMMJ125715.9$+$215432 &  15300216 &&\\ 
H2XMMJ000100.2$-$250459 &   26900002$^a$&  H2XMMJ125712.2$+$215358 &  15300219 &&\\ 
H2XMMJ000036.6$-$250105 &   26900022$^a$&  H2XMMJ125640.3$+$215351 &  15300222 &&\\ 
H2XMMJ000031.7$-$245459 &   26900004$^a$&  H2XMMJ125628.1$+$215405 &  15300225 &&\\ 
H2XMMJ000027.7$-$250441 &   26900001$^a$&  H2XMMJ125710.6$+$215353 &  15300226 &&\\ 
H2XMMJ000026.0$-$250648 &   26900012$^a$&  H2XMMJ125650.3$+$215329 &  15300230 &&\\ 
H2XMMJ003418.5$-$120809 &   15800012$^a$&  H2XMMJ125659.0$+$215347 &  15300231 &&\\ 
H2XMMJ204420.5$-$104904 &   50900061$^a$&  H2XMMJ125654.1$+$215318 &  15300236 &&\\ 
H2XMMJ204446.4$-$103839 &   50900036$^a$&  H2XMMJ125653.7$+$215125 &  15300240 &&\\ 
H2XMMJ204349.2$-$103746 &   50900031$^a$&  H2XMMJ125633.1$+$215147 &  15300253 &&\\ 
H2XMMJ140144.9$+$025332 &  18350140 &  H2XMMJ110522.5$+$381103 &  42100102 &&\\ 
H2XMMJ140144.8$+$024845 &  18350227 &  H2XMMJ110522.0$+$381401 &  42100058 &&\\ 
H2XMMJ140139.6$+$025722 &  18350069 &  H2XMMJ110517.9$+$381051 &  42100106 &&\\ 
H2XMMJ140137.1$+$024604 &  18350258 &  H2XMMJ110512.3$+$382129 &  42100269 &&\\ 
H2XMMJ140132.1$+$025222 &  18350155 &  H2XMMJ110509.7$+$381253 &  42100073 &&\\ 
H2XMMJ140130.8$+$024532 &  18350262 &  H2XMMJ110449.2$+$381810 &  42100302 &&\\ 
H2XMMJ140127.7$+$025607 &  18350095 &  H2XMMJ110447.6$+$380407 &  42100197 &&\\ 
H2XMMJ140125.3$+$024620 &  18350256 &  H2XMMJ110444.2$+$381449 &  42100039 &&\\ 
H2XMMJ140117.5$+$024349 &  18350279 &  H2XMMJ110438.2$+$382500 &  42100243 &&\\ 
H2XMMJ140115.0$+$024821 &  18350240 &  H2XMMJ110437.8$+$382304 &  42100258 &&\\ 
H2XMMJ140109.9$+$024339 &  18350281 &  H2XMMJ110435.2$+$382139 &  42100266 &&\\ 
H2XMMJ140109.0$+$025651 &  18350084 &  H2XMMJ110431.5$+$380305 &  42100207 &&\\ 
H2XMMJ140057.3$+$023942 &  18350033 &  H2XMMJ110424.9$+$380024 &  42100229 &&\\ 
H2XMMJ140053.1$+$024150 &  18350232 &  H2XMMJ110420.8$+$380443 &  42100189 &&\\ 
H2XMMJ140053.1$+$030104 &  18350034 &  H2XMMJ110418.2$+$382047 &  42100272 &&\\ 
H2XMMJ140051.1$+$025906 &  18350048 &  H2XMMJ110416.2$+$380241 &  42100212 &&\\ 
H2XMMJ140049.1$+$025850 &  18350052 &  H2XMMJ110414.7$+$380714 &  42100148 &&\\ 
H2XMMJ140040.9$+$025353 &  18350135 &  H2XMMJ110413.2$+$382203 &  42100264 &&\\ 
H2XMMJ140038.7$+$024322 &  18350283 &  H2XMMJ110402.8$+$375950 &  42100237 &&\\ 
H2XMMJ140033.3$+$025810 &  18350057 &  H2XMMJ110345.8$+$380546 &  42100170 &&\\ 
H2XMMJ140033.0$+$025740 &  18350061 &  H2XMMJ110343.6$+$381349 &  42100055 &&\\ 
H2XMMJ140019.3$+$025638 &  18350080 &  H2XMMJ110339.9$+$380014 &  42100231 &&\\ 
H2XMMJ133807.5$+$242412 &  13390094 &  H2XMMJ110330.2$+$381607 &  42100011 &&\\ 
H2XMMJ133712.7$+$243251 &  13390015 &  H2XMMJ110325.8$+$381212 &  42100088 &&\\ 
H2XMMJ133643.1$+$242646 &  13390060 &  H2XMMJ110322.6$+$380945 &  42100116 &&\\ 
H2XMMJ133630.2$+$242625 &  13390064 &  H2XMMJ110318.6$+$381545 &  42100025 &&\\ 
H2XMMJ133702.2$+$242434 &  13390085 &  H2XMMJ110317.1$+$381336 &  42100066 &&\\ 
H2XMMJ133730.6$+$242306 &  13390103 &  H2XMMJ005019.4$-$515532 &  16274019 &&\\ 
H2XMMJ133717.9$+$242148 &  13390112 &  H2XMMJ005026.3$-$515929 &  16274055 &&\\ 
H2XMMJ133649.4$+$242004 &  13390129 & H2XMMJ005009.4$-$515933 &  16274057 &&\\ 
\hline   
\end{tabular}   
To help the comparison with previous HELLAS2XMM publications, we report the IAU names for the HELLAS2XMM 
source sample and the previously published ones. \newline 
$^a$ Sources belonging to the HELLAS2XMM 1dF sample. 
 
\end{table*}

\end{document}